\documentclass[pdflatex,sn-mathphys-num]{sn-jnl}

\usepackage{array}
\usepackage{pdflscape} 
\usepackage{lscape}
\usepackage{graphicx}%
\usepackage{multirow}%
\usepackage{amsmath,amssymb,amsfonts}%
\usepackage{amsthm}%
\usepackage{mathrsfs}%
\usepackage[title]{appendix}%
\usepackage{xcolor}%
\usepackage{textcomp}%
\usepackage{manyfoot}%
\usepackage{booktabs}%
\usepackage{algorithm}%
\usepackage{algorithmicx}%
\usepackage{algpseudocode}%
\usepackage{listings}%

\theoremstyle{thmstyleone}%
\theoremstyle{thmstyletwo}%

\theoremstyle{thmstylethree}%

\begin{document}

\title[Article Title]{A Comprehensive Review of AI-based Intelligent Tutoring Systems: Applications and Challenges}


\author*[1]{\fnm{Meriem} \sur{Zerkouk}}\email{meriem.zerkouk@teluq.ca}

\author[1]{\fnm{Miloud} \sur{Mihoubi}}\email{miloud.mihoubi@teluq.ca}

\author[1]{\fnm{Belkacem} \sur{Chikhaoui}}\email{belkacem.chikhaoui@teluq.ca}

\affil*[1]{\orgdiv{Artificial Intelligence Institute (I2A)}, \orgname{Université TÉLUQ}, \orgaddress{\city{Quebec}, \state{QC}, \country{Canada}}}


\abstract{AI-based Intelligent Tutoring Systems (ITS) have significant potential to transform teaching and learning. As efforts continue to design, develop, and integrate ITS into educational contexts, mixed results about their effectiveness have emerged. This paper provides a comprehensive review to understand how ITS operate in real educational settings and to identify the associated challenges in their application and evaluation. We use a systematic literature review method to analyze numerous qualified studies published from 2010 to 2025, examining domains such as pedagogical strategies, NLP, adaptive learning, student modeling, and domain-specific applications of ITS. The results reveal a complex landscape regarding the effectiveness of ITS, highlighting both advancements and persistent challenges. The study also identifies a need for greater scientific rigor in experimental design and data analysis. Based on these findings, suggestions for future research and practical implications are proposed. }
\maketitle


\section{Introduction}
Intelliigent Tutoring Systems (ITS) \cite{Makkubhai2023EnhancingEP} \cite{Abbas2023RoleOA} have transformed education, with studies showing that they can improve student performance by 20\%. These advanced computer programs mimic human tutors by offering personalized instruction and feedback tailored to each student's needs. This ability to adjust to individual learning styles makes them valuable in both traditional classrooms and online learning platforms.  ITS have evolved from simple repetitive drill programs to sophisticated systems capable of natural language interactions and adaptive learning.

Although individual human tutoring has demonstrated a 98\% improvement in student performance \cite{Thomas2023ImprovingSL}, its large-scale application is limited by high costs and scalability issues. ITS offer a more accessible and cost-effective alternative, providing personalized instruction without the constraints of human resources. While these systems do not yet fully replicate the nuanced interactions of human tutors, advances in artificial intelligence are gradually bridging this gap. To maximize their effectiveness, ITS must adapt to diverse educational contexts, support multiple languages, and provide contextually relevant content \cite{Ou2024TransformingET}. This flexibility increases their utility as educational tools. Moreover, ITS promote student autonomy and self-regulated learning by providing consistent, personalized feedback, which enhances critical thinking skills.\\

AI-driven ITS also improve accessibility and inclusivity, particularly for students with special needs, by offering personalized support and differentiated instruction. However, their deployment raises ethical concerns, especially regarding data protection and potential biases. Therefore, ensuring robust data protection and algorithmic transparency is essential to maintain trust and safeguard sensitive information. 

Integrating NLP in ITS enhances their ability to engage in meaningful interactions, creating personalized learning pathways that evolve with student progress. These features make ITS a cost-effective and scalable solution compared to traditional educational methods. However, evaluating the effectiveness of ITS in natural educational environments poses methodological challenges, particularly regarding the design of robust social experiments and the measurement of long-term outcomes

The main objective of this review is to highlight the potential and impact of ITS in education by examining the latest models, trends, and techniques in artificial intelligence applied to education. The following specific objectives are directly derived from this primary objective:
\begin{itemize}

\item Identify and classify the unique features of AI-based ITS that distinguish them from traditional tutoring methods.
\item Study the pedagogical approaches and support resources used in AI-based ITS to enhance adaptive learning and personalization.
\item Analyze the integration of ML algorithms and NLP in ITS to improve student interaction, assessment, and overall learning outcomes.
\item Examine student modeling and assessment functionalities within ITS, focusing on their contribution to personalized learning.
\item Evaluate the methods used to measure the pedagogical effectiveness and user satisfaction of AI-based ITS.
\item Study the application and impact of AI-based ITS in specific educational domains.
\item Identify and analyze emerging trends and future technological developments likely to influence the evolution of AI-based ITS.
\item Discuss the challenges faced by ITS in industrial environments.

\end{itemize}
In the following sections, we review the relevant literature on ITS, exploring their definition and architecture in section two. In section three, we detail the systematic review process, starting with the methodology, which includes planning the review, conducting the review, and reporting the review. Section four presents the results and provides a comprehensive discussion of our findings. This is followed by an in-depth exploration of the research questions overview in section five, and we conclude by addressing the challenges and future directions of AI-based ITS.

\section{Intelligent Tutoring Systems (ITS)}
This section covers the foundational aspects of ITS, including precise definitions and architectural details.

\subsection{Definition and architecture}
ITS  are advanced educational software designed to provide individualized instruction and personalized feedback to learners \cite{Kurni2023}. By mimicking the role of human tutors, ITS adapt teaching strategies to the specific needs of students, often without direct human intervention. The primary goal of ITS is to enhance learning by making it meaningful and effective, thereby offering high-quality education accessible to all learners. In modern education, ITS play a crucial role by addressing the diverse needs of students and providing tailored educational experiences. The potential impact of ITS on learning outcomes is significant, as they can help bridge gaps in understanding, offer real-time feedback, and adapt to various learning paces. This flexibility makes ITS particularly valuable in distance learning and large classrooms, where personalized attention from human tutors may be limited.

\begin{figure}[!t]
\centering
\includegraphics[width=0.51\textwidth]{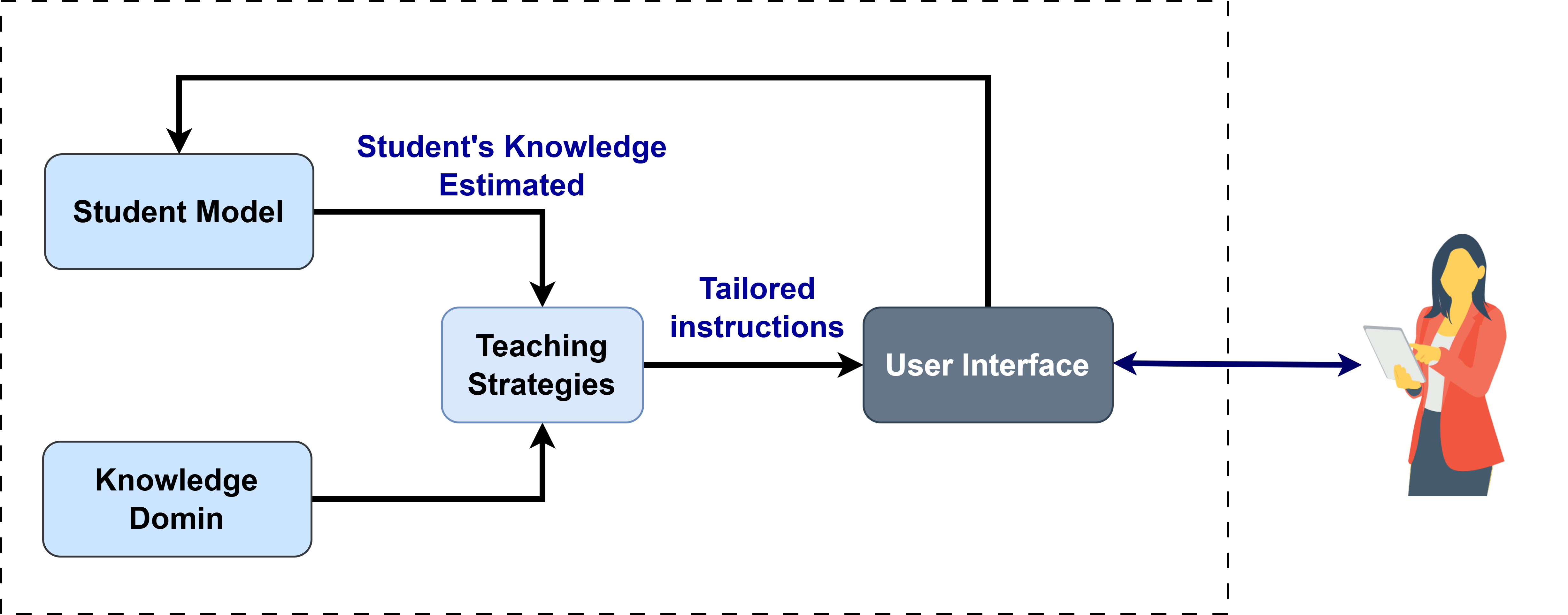}
\caption{ITS Architecture \cite{Qwaider2018ExcelIT}.}
\label{architecture}
\end{figure}

In \cite{Nkambou2010AdvancesII}\cite{Qwaider2018ExcelIT}, the authors describe the four-component architecture, as illustrated in Figure \ref{architecture}, a widely accepted framework for defining ITS \cite{Fang2018AMO}. This architecture typically includes the Domain Model, the Student Model, the Tutor Model, and the User Interface. The Domain Model encapsulates the concepts, rules, and problem-solving strategies pertinent to a specific learning domain. It serves as an expert knowledge base, providing the foundation for assessing student performance and identifying errors. The Domain Model includes a comprehensive representation of the subject matter, allowing the ITS to compare a student's input against expert knowledge to provide accurate feedback. The Student Model continuously monitors learners' progress, adapting to their cognitive and affective states. This component tracks students' understanding, skills, and misconceptions by comparing their actions and responses to the Domain Model. Methods such as Bayesian network\cite{Santhi2013ReviewOI}, ML algorithms \cite{Lin2023ArtificialII}, and data mining techniques \cite{Chen2019ResearchOI} are often employed to refine the model's accuracy in predicting student needs and learning paths.

The Tutor Model determines the instructional strategies and interventions necessary to facilitate learning. It decides when and how to intervene, what content to present, and how to deliver it. The Tutor Model  utilizes pedagogical rules and decision-making algorithms to customize feedback and guidance, ensuring that interventions are timely and appropriate for the learner's current state. 

The User Interface (UI) facilitates student and system communication. It is designed with principles of effective human-computer interaction to ensure that the educational content is accessible and engaging. The UI integrates the necessary tools for dialogue generation, visual representations, and interaction, making the learning experience intuitive and user-friendly.

\section{Methodology}
This study follows the recommendations of Kitchenham et al. \cite{Budgen2006PerformingSL} for conducting a systematic literature review organized into three main stages. The first stage, planning, defines the needs of the review and formulates specific research questions. The second stage, conducting the review, involves the systematic search for relevant articles, the application of inclusion and exclusion criteria, and systematically extracting data from selected studies. The third and final stage, reporting the review, concerns the structured presentation of the results, aligning with the reporting elements recommended for systematic reviews and meta-analyses PRISMA established by Moher et al. \cite{Page2020TheP2}. This methodology ensures comprehensive coverage of publications on ITS, assessing their effectiveness and innovation through key performance indicators (KPIs) while providing the reproducibility and transparency of the review process. The selection process followed these steps: 37,617 articles were initially identified, 9,485 were eliminated after deduplication, 2,112 were excluded after screening titles and abstracts, and finally, 127 articles were included for full review, supplemented by 26 reports from specialized web sources.

\subsection{Planning the Review}
\begin{figure}[!h]
\centering
-\includegraphics[width=0.8\textwidth]{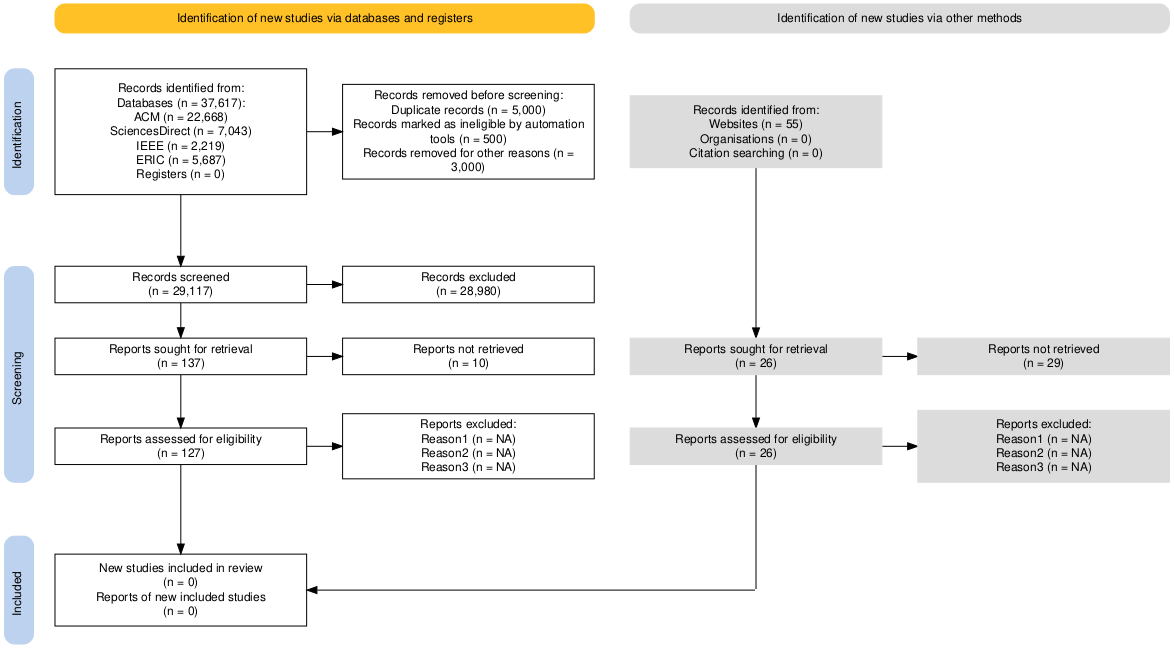} 
\caption{PRISMA Flow Diagram of the study selection process}
\label{fig:PRISMA}
\end{figure}

A search for studies was carried out in several well-known educational technology databases, including Web of Science, Scopus, IEEE Xplore, and Springer. The search included relevant articles published up to 2025, as illustrated in Figure \ref{fig:PRISMA}. There has been a marked increase in the number of articles in recent years. The study included various types of publications, such as research articles, review papers, and book reviews. A combination of search terms, described in Table \ref{tab:search_terms}, was used to address the complexity of the issue, specifically focusing on ITS. This process yielded a total of 37,617 articles, which were then filtered according to the inclusion and exclusion criteria presented in Table \ref{tab:inclusion_exclusion_criteria}. The temporal distribution of the selected publications spans primarily from 2010 to 2025, with a significant increase observed starting in 2015 and a peak between 2019 and 2023 a period marked by the rise of generative AI technologies applied to ITS.

\begin{table}[h!]
\centering
\caption{Search terms}
\begin{tabular}{|>{\centering\arraybackslash}m{1.8cm}|>{\centering\arraybackslash}m{1.8cm}|>{\centering\arraybackslash}m{1.8cm}|>{\centering\arraybackslash}m{2cm}|}
\hline
A & B & C & D\\ \hline
intelligent* & tutoring & system & experiment* \\ \hline
adaptive & instruction & software & trial \\ \hline
personalized & education & application & evaluat* \\ \hline
AI & learning & program & social experiment \\ \hline
machine & mentoring & platform & assessment \\ \hline
\end{tabular}
\label{tab:search_terms}
\end{table}

The collected papers were then filtered according to the inclusion and exclusion criteria presented in Table \ref{tab:inclusion_exclusion_criteria}.

\begin{table}[h!]
\centering
\caption{Inclusion and Exclusion Criteria}
\begin{tabular}{|p{4cm}|p{4cm}|}
\hline
Inclusion & Exclusion \\ \hline
Focused on the evaluation of educational effectiveness of ITS. & Focused on the evaluation of ability or technical performance of ITS. \\ \hline
Studies conducted in real learning environments. & Studies conducted outside of real learning environments (e.g., technology development lab). \\ \hline
Empirical studies with data or evaluation results. & Proposals or prototypes of ITS without evaluation results. \\ \hline
Articles written in English. & Articles not written in English. \\ \hline
Validation duration of 6 weeks or more. & Validation duration less than 6 weeks. \\ \hline
Sample size of 100 participants or more. & Sample size less than 100. \\ \hline
Peer-reviewed papers. & Papers not peer-reviewed. \\ \hline
\end{tabular}
\label{tab:inclusion_exclusion_criteria}
\end{table}

To visually represent the selection process, we utilized the PRISMA flow diagram, as shown in Figure \ref{fig:PRISMA}. This diagram outlines the flow of information through the different phases of the systematic review, including identification, screening, eligibility, and inclusion of studies. 
To achieve our objective, we started by planning this scoping review \cite{Tricco2016ASR}. The initial task involved defining the research questions (RQs); we aimed to explore specific questions detailed in the following sections of this paper.

The following research questions are formulated to address the main objective and its associated sub-goals:

\begin{itemize}
\item RQ1: What are the key features of AI-based ITS that distinguish them from traditional tutoring methods?
\item RQ2: What pedagogical strategies and support resources are employed by ITS to enhance adaptive learning and personalization?
\item RQ3: How can ML and NLP be integrated into ITS to improve student interaction and assessment?
\item RQ4: How do student modeling and assessment features in ITS contribute to personalized learning?
\item RQ5: How are AI-based ITS evaluated regarding pedagogical effectiveness and user satisfaction?
\item RQ6: What progress have AI-based ITS made in specific applications areas such as mathematics, science, or languages?
\item RQ7: What are the emerging trends and future technologies likely to influence the development of AI-based ITS?
\item RQ8: What advancements have AI-based Industrial ITS made in specific application areas? 
\end{itemize}

\subsection{Conducting the Review}
The process of applying the inclusion/exclusion criteria was carried out with the utmost objectivity. Multiple independent reviewers screened the titles and abstracts of the identified articles. Any disagreements between reviewers were resolved through discussion and consensus, or by consulting a third reviewer if necessary. After applying the inclusion/exclusion criteria, 127 articles remained for full text review. To ensure the rigor of article selection, the inter-rater agreement coefficient was calculated on a random sample of 200 articles, yielding a Cohen's Kappa score of 0.82, indicating an excellent level of agreement

The data extraction process involved systematically extracting relevant information from each selected article. This included details such as: study objectives and research questions, methodology and experimental design, sample size and participant demographics, key findings and results, and any reported limitations or biases.

\subsection{Reporting the Review}
The reporting stage involved structuring the presentation of results according to PRISMA guidelines to ensure clarity, transparency, and reproducibility. By analyzing the selected articles, this review provides an overview of the current implementation of ITS and the integration of AI. It explores how AI supports ITS and evaluates their effectiveness. Additionally, the review addresses challenges and considerations associated with implementing ITS, such as assessing AI effectiveness and managing technical difficulties. The findings are presented in a structured manner, adhering to PRISMA guidelines.

\section{Review Findings}

This section presents the addressed research questions and the key findings. 

\textbf{RQ1. What are the key features of AI-based ITS that distinguish them from traditional tutoring methods?} \\

This review explores the integration of key features in ITS, such as adaptive learning, personalization, learner modeling, real-time feedback, and data-driven insights. It examines how these systems customize learning experiences to meet individual student needs and preferences, as well as their impact on learner engagement and educational outcomes. Numerous ITS have been developed to support adaptive and personalized learning (PL), and this review identifies systems that effectively incorporate these features. Additionally, it provides an in-depth analysis of the benefits of ITS in delivering personalized and adaptive learning.

AI-powered ITS distinguish themselves from traditional tutoring methods through unique features like dynamic curricula, real-time feedback, scalability, enhanced engagement, and seamless integration with other technologies. These advantages result in a more efficient, effective, and engaging learning experience for students. To elucidate the distinguishing features of AI-based ITS, we consider the following aspects:\\

Personalized Learning: ITS represent an advanced approach to personalized learning by leveraging sophisticated algorithms to tailor educational experiences to individual learners. Effective personalization \cite{Bellarhmouch2022APA} in education requires a comprehensive understanding of the learner and the design of meaningful and relevant learning tasks. Given learners' diverse needs and characteristics, creating an environment that adapts to these variations is inherently complex. The learner model, a critical component of adaptive environments, encapsulates the learner's profile and serves as the foundation for personalization. By accurately modeling the learner's needs and characteristics, ITS can adjust the learning environment dynamically, resulting in a more effective and efficient educational experience.

In AI-based ITS, personalized learning involves customizing educational content and pacing to match each student's unique abilities and requirements. This targeted approach addresses individual learning gaps and leverages student strengths, leading to more effective knowledge acquisition and retention. For example, specific algorithms like Bayesian networks \cite{Santhi2013ReviewOI}, decision trees \cite{Crockett2017OnPL}, and deep learning models are employed to continuously update and refine the learner model \cite{Jiang2024BeyondAL}, allowing for real-time adaptation of instructional strategies. Empirical studies consistently show that students in personalized learning environments outperform their peers in standardized settings \cite{BalasoAIEnhancedEP}, mainly due to the tailored support they receive, which aligns with their specific learning needs.

Kochmar et al. \cite{Kochmar2020AutomatedPF} demonstrated this with the Korbit learning platform, a large-scale, dialogue-based ITS. This model meets personalization by generating automated, data-driven personalized feedback using advanced ML and NLP techniques. The system dynamically adapts its instructional content based on the student's individual needs, providing personalized hints, explanations, and mathematical guidance. The system can tailor educational interventions according to the learner's profile and their zone of proximal development (ZPD), optimizing learning outcomes and enhancing student engagement by employing models such as neural networks and reinforcement learning algorithms.

Bernacki et al. \cite{Bernacki2021ASR} systematically explored and clarified the multifaceted landscape of personalized learning (PL) by examining who is responsible for personalizing learning, what elements are personalized, how personalization is implemented, and the purposes it serves. The article underscores the variability in definitions and implementations of PL across different educational contexts, complicating the assessment of how effectively PL designs leverage student characteristics to enhance learning outcomes.

Laak et al. \cite{Laak2024AIAP} implemented a dynamic student modeling approach combined with the zone of proximal development (ZPD) concept. In their ITS, the student model is continuously updated based on the learner's interactions, capturing cognitive states, knowledge levels, and affective states. By aligning instructional content within the learner's ZPD, the ITS ensures that tasks are both challenging and achievable, optimizing learning outcomes and fostering deeper engagement. This approach integrates insights from cognitive science and educational psychology, creating a more effective and personalized learning environment that directly addresses the limitations of traditional, one-size-fits-all tutoring methods.

Shah et al. \cite{Shah2024AIPoweredPL} and \cite{Wang2025LLMpoweredMF} explored the use of LLMs to enhance personalization within ITS by enabling more natural and dynamic interactions between the system and learners. LLMs, such as those in the GPT series, are trained on vast datasets and are incorporated into ITS to understand and generate human-like responses. This capability allows ITS to provide individualized feedback, clarify concepts, answer questions, and engage in dialog-based tutoring sessions that adapt to the learner's needs. By leveraging the advanced capabilities of LLMs, these systems overcome traditional tutoring's limitations, such as scalability and consistency, offering a more flexible and responsive learning experience.

Adaptive Learning :
Systems that include ITS  \cite{Brusilovsky2007UserMF} \cite{Paredes2004AMA} and Adaptive ITS (AITS), dynamically adjust learning paths, pacing, and content to meet each student’s unique needs and preferences. By integrating adaptive features, these systems ensure that learning experiences are efficient and personalized. Adaptive mechanisms, such as real-time assessments of student progress, are employed to adjust instructional content and difficulty levels, maintaining an optimal challenge that prevents boredom from overly simple material and frustration from challenging content \cite{Ermit2020DesignFO}. This adaptability keeps students engaged and motivated, especially in diverse environments where students have varying backgrounds and levels of prior knowledge. Furthermore, the long-term effects on student motivation and self-regulated learning are significant, as these systems foster sustained interest and autonomy by tailoring materials to each student's progression level \cite{Walkington2013UsingAL} \cite{AlNakhal2017AdaptiveIT}.

Phobun et al. \cite{Phobun2010AdaptiveIT} explained that AITS seamlessly integrate ITS with Adaptive Hypermedia (AH) to create personalized and dynamic learning experiences. AITS are composed of an expert model that adapts educational content and delivery based on a learner's current knowledge, learning style, and progress. By leveraging adaptive presentation and navigation, AITS ensure that each learner receives the most relevant material optimally. This real-time adaptability, grounded in a deep understanding of learner interactions, enhances theoretical and practical learning, making AITS a powerful tool for personalized education across various domains.

To personalize learning, Liu et al. \cite{Liu2024TheDO} employed adaptive prompts in ITS grounded in educational theories. These prompts decompose complex problems, foster critical thinking, and offer tailored cognitive support. By dynamically adjusting to student performance, ITS enhance problem-solving efficiency and deepen understanding, ensuring a highly personalized and practical learning experience.\\

Learner Modeling: Learner models play a critical role in ITS by representing essential user-specific information. As discussed by \cite{Brusilovsky2007UserMF}, user models can include the overlay approach, where user knowledge is mapped as a subset of expert domain knowledge, and the uncertainty-based approach, which uses probabilistic methods to manage the uncertainties in understanding user behaviors and preferences. This enables ITS to finely tune the content, navigation, and interface to each user's needs. Paredes et al. \cite{Paredes2004AMA} introduce a theoretical framework for dynamically modeling student learning styles, combining explicit data from questionnaires and implicit data from ongoing course interactions. Learner modeling involves creating detailed representations of student’s cognitive and emotional states. These models enable ITS to provide targeted interventions, such as adjusting the complexity of tasks or offering additional emotional support when frustration is detected. By accurately identifying these challenges, ITS can help students overcome specific obstacles, leading to a more supportive and effective learning experience. Enhanced learner modeling can also include emotional \cite{Xu2018LearningEE} and cognitive state recognition \cite{ZataranCabada2014EmotionRI} to provide real-time, context-sensitive responses, thereby improving student engagement and well-being. Kumar et al. \cite{Kumar2019AnAF} argue that accurately modeling and understanding the learner's characteristics and behaviors allows ITS to deliver a highly personalized and effective educational experience. This adaptivity is achieved through continuous assessment and real-time adjustments based on the learner's interactions and progress.\\

Recommender Systems: The application of recommender systems within Technology Enhanced Learning (TEL) is crucial, as explored by \cite{Manouselis2012RecommenderSF}. These systems help find relevant learning resources, enhancing the learning experience by providing personalized content recommendations. Recommender systems in ITS suggest personalized learning resources, enhancing educational inclusivity by catering to diverse learning styles and cultural backgrounds. However, there is a risk of reinforcing biases if the recommendation algorithms are not carefully designed. To address this issue, it is crucial to ensure that the data used for training these systems is both diverse and representative. Additionally, regular audits and adjustments of the recommendation criteria help maintain fairness and inclusivity. Ensuring diversity and inclusivity in these recommendations can prevent content bias and support a broader educational scope. Muangprathub et al.\cite{Muangprathub2020LearningRW} focused on developing a learning recommendation component within ITS that can dynamically predict and adapt to individual learners' styles. The primary focus is on creating an adaptive algorithm and an improved knowledge base that enables personalized learning by providing tailored content recommendations.\\

Real-Time Feedback and Assessment: One of the most important advantages of AI-based ITS is the ability to provide immediate feedback \cite{Shute2007FocusOF} \cite{Eduaide2024}. These systems assess student responses in real-time and offer corrective feedback \cite{Modi2022IncorporatingFT}, allowing students to learn from their mistakes quickly. Effective student feedback is a critical tool for improving the quality of education by enhancing instructional methods, refining curricula, and ultimately improving learning experiences and outcomes \cite{Weaver2024ThePO}. The impact of feedback depends on how it is processed and utilized by institutions. Positive feedback can boost student confidence, while negative feedback, if not delivered thoughtfully, can lead to anxiety or decreased motivation. To minimize these adverse effects, ITS should employ strategies such as constructive criticism \cite{Cutumisu2017AssessingWS}, emphasizing effort over inherent ability, and providing balanced feedback highlighting strengths and improvement areas. Careful design is essential to avoid reinforcing negative self-perceptions or anxiety, particularly in formative assessments. This strategic use of feedback ensures continuous improvement in education, making the learning process more responsive to student needs and more effective overall \cite{Abdi2023StudentsFA}. LLMs enhance feedback in ITS by providing personalized, real-time responses that adapt to individual learner needs. These models can offer nuanced and empathetic feedback, improving the learning experience. However, for LLMs to be truly effective, their use must be grounded in solid theoretical frameworks and validated through empirical research to ensure that their feedback effectively supports student learning.\\

Data-Driven Insights and Explainable AI (XAI): The integration of data-driven insights \cite{Koedinger2013NewPF} in ITS significantly enhances the transparency and explainability of the system's decision-making processes. ITS can clarify these processes by employing mechanisms such as XAI, thereby building user trust. Data-driven insights contribute to transparency by offering clear, evidence-based explanations for instructional decisions. Implementing XAI in ITS involves using models that provide understandable rationales for their recommendations and actions. Best practices include offering users accessible explanations of how decisions are made and ensuring that these systems are interpretable by educators, students, and parents, thus fostering trust and accountability.\\  Conati et al. \cite{Conati2019TowardPX} propose that personalizing explanations in XAI within ITS can significantly enhance students' trust, perceived usefulness, and learning outcomes. Their study shows that tailoring explanations to individual learner characteristics, such as 'Need for Cognition,' 'Conscientiousness,' and 'Reading Proficiency,' can make AI-driven educational support more effective and engaging, underscoring the potential value of personalized XAI in improving educational experiences.
Karpouzis et al. \cite{Karpouzis2024ExplainableAI} integrate XAI in educational technologies, as demonstrated by the iRead project, plays a crucial role in enhancing learning outcomes, student engagement, and teacher acceptance by providing transparency and adaptability. The project illustrates how XAI not only supports personalized learning and manages classroom diversity but also alleviates teacher concerns, leading to more effective integration of AI into educational settings.
Ali et al.\cite{Ali2023ExplainableAI} emphasize the need for tailored explanations to suit different user types and presents a hierarchical categorization of XAI techniques, including data, model, and post-hoc explainability, along with the evaluation of explanations. This study is especially relevant for those applying XAI in domains such as ITS, where explainability is critical for user trust and comprehension.

Gamification and Engagement: Gamification in ITS involves integrating game like elements, such as levels, points, and progress bars, into the learning process to enhance engagement and motivation \cite{Alsawaier2018TheEO}. This approach aims to overcome challenges like boredom, lack of interest, or monotony when using traditional ITS \cite{Andrade2016TheBA}. By incorporating these elements, gamification can make learning more interactive and enjoyable, potentially leading to improved learning outcomes. The combination of ITS and gamification, referred to as ITS+G, has shown promise in both STEM and non-STEM subjects, offering a more dynamic and responsive educational experience \cite{Ramadhan2023ITSGamification}.\\
Huang et al \cite{Huang2020TheIO} highlight that integrating gamification elements, such as badges and leaderboards, into ITS can enhance student engagement and motivation, potentially leading to improved learning outcomes. However, the effectiveness of gamification varies depending on the design elements used, the educational context, and the characteristics of the learners.\\

Affective Intelligent Tutoring Systems (AITS):  leverage advancements in affective computing to create more empathetic and responsive learning environments \cite{FernndezHerrero2024EvaluatingRA}. These systems can adapt their instructional strategies in real-time by detecting and interpreting students' emotional states, such as frustration, confusion, or disengagement. For example, when a student exhibits frustration, an Affective ITS might provide additional hints, adjust the difficulty level, or offer encouraging feedback to alleviate the negative emotion and re-engage the learner. This emotional awareness enhances the tutoring process's effectiveness and contributes to a more supportive and personalized learning experience. Affective ITS thus hold significant potential in fostering a positive learning atmosphere, leading to improved learning outcomes and increased student motivation across various educational contexts. \cite{Hasan2020TheTF}.\\
The evaluation of AI-based ITS, as presented in Table~\ref{tab:key_features_ITS_with_KPIs_objective}, offers a comprehensive overview of various ITS implementations that integrate advanced AI technologies, including Personalized Learning, Real-Time Feedback, Adaptive Learning, and Data-Driven Insights. These systems employ ML, NLP, and XAI to enhance the learning experience through improved student engagement, timely feedback, and adaptive instructional strategies. The table highlights numerous studies and their corresponding objectives, demonstrating how AI enables ITS to model learner behavior, deliver personalized feedback, and recommend individualized learning paths. Leveraging these technologies allows ITS to achieve greater accuracy in adaptive learning, increased engagement via gamification, and emotional support through affective computing. While many studies, such as \cite{Bellarhmouch2022APA, Crockett2017OnPL, Jiang2024BeyondAL}, underscore the effectiveness of AI-driven ITS in delivering real time personalization and adaptive instruction, their evaluation methodologies differ significantly. Some rely on subjective engagement indicators, whereas others use standardized performance assessments. This methodological inconsistency complicates direct comparisons and limits the generalizability of findings. Moreover, only a few works such as \cite{BalasoAIEnhancedEP} investigate the long term impact or the adaptability of ITS across diverse learner profiles and educational settings.  This lack of longitudinal and cross contextual validation suggests that the observed benefits may be context-dependent rather than universally generalizable. Additionally, the variability in how learner models are constructed and updated ranging from probabilistic frameworks to deep learning approaches raises critical concerns about transparency, fairness, and interpretability in AI mediated education.
To reinforce the empirical foundation of AI-based ITS, future research should adopt unified key performance indicators (KPIs), as partially illustrated in Table~\ref{tab:key_features_ITS_with_KPIs_objective}, and rigorously assess learner diversity, fairness in recommender systems, and long term learner engagement. Such efforts would support more consistent benchmarking and promote the scalable and equitable deployment of ITS in real world educational environments. Nevertheless, recent scholarship cautions that the widespread adoption of these systems remains constrained by unresolved challenges, including student data privacy, algorithmic bias, and issues of academic integrity \cite{Gao2025Agent4EduGL}.\\


\textbf{RQ2. What pedagogical strategies and support resources are employed by ITS to enhance adaptive learning and personalization?}\\

AI techniques play a pivotal role in implementing pedagogical strategies that deliver personalized instruction for ITS \cite{Lin2023ArtificialII}. These systems employ various methods to create adaptive, personalized learning experiences. Expert systems utilize predefined rules and domain knowledge to guide instruction, while case-based reasoning systems solve new problems by referencing past cases. Cognitive tutors models support students’ cognitive processes, enabling ITS to mimic the adaptive capabilities of human tutors and provide effective and individualized education.

Pedagogical Strategies in ITS refer to the methods and techniques used to deliver instruction and facilitate learning. These strategies are designed to adapt to diverse learning styles, preferences, and paces of students, with the goal of creating a learning environment that is effective, engaging, and motivating. Key aspects include personalization, adaptability, feedback and assessment, and engagement. These strategies are often implemented using rule-based systems, expert systems, case-based reasoning, and cognitive tutors, leveraging AI techniques such as knowledge representation and reasoning capabilities.

Rule-based and Expert Systems:
Rule-based and expert systems guide instruction and provide feedback within ITS by utilizing predefined rules and knowledge bases. These systems emulate the decision-making processes of human experts in specific domains. Key features include using a set of if-then rules to determine appropriate instructional actions, incorporating domain-specific knowledge to provide targeted guidance, and adapting instruction based on learner responses and performance\cite{Frasson1992FromES}
\cite{Hafidi2013DesignAE}
\cite{Zarandi2012AFE}.

Some of the earliest and most influential rule-based ITS include GUIDON \cite{Clancey1987KnowledgebasedTT}, developed by William Clancey in the 1980s for teaching medical diagnosis, and the LISP Tutor \cite{Clancey1987KnowledgebasedTT}, developed by John Anderson and colleagues at Carnegie Mellon University. Other notable examples include the Andes Physics Tutor, SQL-Tutor, and AutoTutor\cite{Graesser2006AutoTutorAC}, each employing a rule-based approach to provide individualized problem-solving support in their respective domains.

Expert systems in pedagogical strategies represent and reason with pedagogical knowledge to emulate the decision-making process of human experts, such as experienced teachers or tutors. They can also incorporate ML techniques to refine and update the knowledge base based on new data or experiences.
For instance, Isaiah et al.\cite{Isaiah2015RuleBM} investigate the use of rule-based student models within a multi-agent ITS framework. Their system is designed to understand and assess student comprehension by providing immediate feedback and tailored problems, improving learning outcomes across diverse domains. An expert system \cite{Frasson1992FromES} in ITS is a component that simulates the decision-making ability of a human expert to provide personalized instruction and feedback. It uses a knowledge base and inference engine to understand complex learning scenarios, assess students' performance, and offer tailored guidance. The expert system enhances ITS by enabling it to adapt to individual learners' needs, troubleshoot learning difficulties, and recommend learning paths, thereby creating a more interactive and effective learning environment. Bradc and al. \cite{Bradc2022DesignOA} simulate human expert decision-making to personalize learning. It identifies students' needs, such as their input knowledge and sensory learning preferences, and adapts the learning experience accordingly.



\begin{sidewaystable}[h!] 
\centering
\caption{Key Features of AI-Based ITS with KPIs, Objectives, Findings, and Impact}
\scriptsize 
\begin{tabular}{|p{0.4cm}|p{0.4cm}|p{0.4cm}|p{0.4cm}|p{0.4cm}|p{0.4cm}|p{0.4cm}|p{0.4cm}|p{0.4cm}|p{0.4cm}|p{0.4cm}|p{0.4cm}|p{0.4cm}|p{2cm}|p{2cm}|p{3cm}|}
\hline
Pap & Per & RTF & Adp & LMd & DDI & Eng & Gam & Rec & Aff & Alg & Adt & Acc & Obj & Fin & Imp \\ \hline
\cite{Kochmar2020AutomatedPF} & X & X &  & X &  &  &  &  &  & X &  &  & Automate personalized & Found improvement & Improved learning \\ \hline
\cite{Bernacki2021ASR} & X &  & X &  &  &  &  &  &  &  & X &  & Systematic exploration of & Identified factors & Provides a framework \\ \hline
\cite{Laak2024AIAP} & X &  & X & X &  &  &  &  &  & X &  & X & Implement dynamic student  & Demonstrated improved learner  & Enhanced engagement and  \\ \hline
\cite{Shah2024AIPoweredPL} & X & X &  & X &  & X &  &  &  &  & X &  & Use of LLMs for personalization & Showed flexibility and scalability of & dsds \\ \hline
\cite{Phobun2010AdaptiveIT} &  & X & X & X &  &  &  &  &  &  &  & X & Integrate Adaptive & Enhanced adaptation & Personalized \\ \hline
\cite{Liu2024TheDO} &  &  & X &  &  & X &  &  &  &  & X &  & Use adaptive prompts  & Increased learner critical  & Enhanced learning  \\ \hline
\cite{Brusilovsky2007UserMF} & X &  & X & X &  &  &  &  &  &  & X &  & Model user knowledge  & Found personalized  & Improved learning outcomes with  \\ \hline
\cite{Paredes2004AMA} & X &  & X & X &  &  &  &  &  &  & X &  & Develop learner models & Identified improvement in & Enhanced adaptation to  \\ \hline
\cite{Shute2007FocusOF} &  & X &  &  &  &  &  &  &  &  & X &  & Provide immediate feedback for student performance & Found real-time feedback & Faster learning \\ \hline
\cite{Koedinger2013NewPF} &  & X &  & X & X &  &  &  &  & X &  & X & Use data-driven s & Increased  & Improved trust and adoption  \\ \hline
\cite{Ramadhan2023ITSGamification} &  &  &  &  &  & X & X &  &  &  &  & X & Combine ITS with & Improved motivation & Higher learner engagement \\ \hline
\cite{Hasan2020TheTF} &  &  & X &  &  & X &  &  & X & X &  & X & Use affective computing for  & Improved emotional awareness  & Higher student motivation  \\ \hline
\cite{Muangprathub2020LearningRW} & X &  & X &  &  &  &  & X &  & X & X &  & Develop a learning  & Improved learning by adapting & Enhanced personalization \\ \hline
\cite{Manouselis2012RecommenderSF} &  &  &  &  &  &  &  & X &  &  &  &  & Explore recommender systems in TEL to find  & Enhanced learning experience  & Promoted educational inclusivity by catering s \\ \hline
\cite{Conati2019TowardPX} & X &  & X & X & X & X &  &  &  & X & X &  & Personalizing XAI to learner characteristics to enhance learning & Personalizing explanations increases trust & Significant improvement in student engagement and trust \\ \hline
\cite{Karpouzis2024ExplainableAI} & X & X & X & X & X & X &  &  & X & X & X & X & Integrate XAI to support transparency in learning & Enhanced  & Effective integration of AI in classrooms \\ \hline
\end{tabular}
\label{tab:key_features_ITS_with_KPIs_objective}
\end{sidewaystable}



This allows ITS to provide individualized study plans and guidance, enhancing the learning process automatically and at scale.
Both rule-based systems and expert systems are valuable for implementing pedagogical strategies in ITS because they allow for the explicit representation and application of domain knowledge and instructional principles. However, these systems can be complex to develop and maintain, especially for domains with a large number of rules or a high degree of complexity in the pedagogical strategies.

Case-based Reasoning (CBR):
CBR systems draw upon a library of past cases or scenarios to provide personalized instruction \cite{Gonzlez2013DesigningIT} \cite{SotoForero2024TheIT}. This approach allows the system to adapt to new situations by referencing similar past experiences. Key features include maintaining a database of previous learning scenarios and outcomes, matching current learner situations with similar past cases \cite{Gonzlez2006ACA} 
\cite{Stottler1999ACR} , and providing tailored instruction based on successful strategies from similar cases \cite{Soh2007IntegratedIC} 
\cite{Masood2017CasebasedRI}.
Notable examples of CBR in ITS include  CBR-Tutor \cite{Reyes2001AgentRA} and CBR-Works \cite{Perner2019CaseBasedR}, each using past cases to guide new learning situations in fields such as software engineering, mathematics, and physics. While effective, the success of CBR systems depends on the quality and relevance of stored cases, and they can become computationally intensive when managing and retrieving relevant cases from large datasets. In \cite{Yan2024ARO},  CBR remains a valuable AI method, but its evolution will require addressing the balance between complexity and interpretability, especially as AI systems become more advanced.

Cognitive Tutors:
Cognitive tutors are based on cognitive models of how students acquire and apply knowledge in a particular domain. These systems provide step-by-step guidance and feedback as learners work through problems\cite{Smalenberger2022TowardAI}
\cite{Sychev2024EducationalMF}. Key features include modeling the cognitive processes involved in problem-solving, providing immediate feedback on each step, and adapting instruction based on the learner's understanding and misconceptions.
Research in cognitive tutors has demonstrated their effectiveness in real-world educational settings\cite{Bb2022UsingSA}, with notable studies exploring their application in mathematics, programming, and other domains. Cognitive tutors are praised for their personalized learning capabilities and real-time feedback, though they can be resource intensive and complex to develop.
Carnegie Learning's Cognitive Tutor \cite{Smalenberger2022TowardAI}, ACT-R Tutors \cite{Anderson1997ACTRAT} , SQL-Tutor \cite{Wang1997SQLTA} , Andes Physics Tutor \cite{VanLehn2010TheAP}, Electronix Tutor \cite{Graesser2018ElectronixTutorAI} .

While cognitive tutors have shown promise, it's important to note that they can be complex and resource-intensive to develop. The creation of accurate cognitive models requires extensive domain expertise and iterative refinement. Additionally, the effectiveness of these systems can vary depending on the specific implementation and the context in which they are used. The evaluation of pedagogical strategies integrated into ITS, as presented in table ~\ref{tab:RQ2}, offers a comprehensive overview of various ITS implementations that incorporate advanced pedagogical and AI technologies. These systems employ a range of methodologies, including rule-based systems, expert systems, CBR, cognitive models, ML techniques, to create adaptive and personalized learning environments. The table highlights diverse studies, each aiming to enhance specific aspects of the ITS.  However, despite the demonstrated potential of these pedagogical strategies, many ITS still struggle with generalizability across diverse learning contexts, raising concerns about their adaptability to heterogeneous learner populations. Moreover, the development and maintenance of systems like expert or cognitive tutors remain resource-intensive and difficult to scale, highlighting a critical need for more sustainable and transferable ITS design approaches.
Because the great majority of studies on pedagogical scaffolds are < 6-week, single-site pilots involving fewer than 100 learners, their long-term durability and classroom-level generalisability remain unproven. In addition to these AI-based approaches, many ITS pedagogical designs implicitly or explicitly draw on well established learning theories. For example, constructivist learning theory supports the design of exploratory environments and problem based learning scenarios, where learners actively construct knowledge through scaffolded challenges  a principle evident in cognitive tutor frameworks and case-based reasoning systems.  Likewise, Self-Regulated Learning (SRL) theory is reflected in ITS features such as learner modeling, formative feedback, and goal setting tools, which foster metacognitive strategies including planning, monitoring, and self reflection. Furthermore, Cognitive Load Theory (CLT) informs ITS design by guiding the adjustment of task complexity and the use of stepwise hints to manage intrinsic and extraneous cognitive load. This ensures that learners allocate more effort to germane processing, thereby enhancing meaningful learning while avoiding cognitive overload.\\

\textbf{RQ3. How are ML and NLP integrated into ITS to improve student interaction and assessment?}\\

ML \cite{Sindi2005AML} and NLP \cite{Lan2024SurveyON} are essential components in the development of ITS \cite{Dzikovska2014BEETLEID}. These technologies enable ITS to provide personalized, adaptive, and engaging learning experiences that closely replicate the capabilities of human tutors \cite{Lin2023ArtificialII}. Through various techniques, such as dialogue systems, automated essay scoring, adaptive learning algorithms, and predictive models, ML and NLP contribute to more effective and tailored learning experiences. The following sections demonstrate how the technologies enhance student interaction and assessment in ITS by enabling dynamic, real-time responses and personalized feedback \cite{Paladines2020ASL}.

Dialogue Systems: 
Dialogue systems in ITS explore NLP techniques to facilitate natural language interactions between students and the tutoring system \cite{Graesser2001IntelligentTS} \cite{Heffernan2000IntelligentTS}. 
Diane et al. \cite{Litman2004ITSPOKEAI} introduced ITSPOKE, an intelligent tutoring spoken dialogue system that combines NLP and speech recognition technologies with intelligent tutoring capabilities. The system was designed to engage students in spoken tutorial dialogues about physics concepts. ITSPOKE contributes to understanding how dialogue systems can be effectively used in educational contexts, particularly for subjects involving conceptual and quantitative elements like physics. It also highlights the potential benefits of multimodal interaction in ITS. Pipatsarun et al. \cite{Phobun2010AdaptiveIT} focus on integrating ITS with Adaptive Hypermedia systems (AH) to create an Adaptive Intelligent Tutoring System (AITS) for e-learning. This integration aims to enhance the adaptation and personalization of instruction based on learners' behavior, learning styles, and current knowledge levels.
Algherairy et al. \cite{Algherairy2023ARO} indicate that GPT and similar LLMs represent a significant advancement in dialogue systems by offering more natural, context-aware responses that closely mimic human interaction.
Feng et  al. \cite{Feng2024CourseAssistPA} highlight the role of GPT and similar LLMs in enhancing dialogue systems for ITS. CourseAssist1, an LLM-based tutoring system, tailors GPT's question-answering capabilities to provide effective, scalable tutoring in computer science education.
 The authors in \cite{Piro2024MyLearningTalkAL} emphasize the GPT and other LLMs in the development of dialogue-based ITS. The introduction of MyLearningTalk (MLT), a web-based ITS powered by LLMs, demonstrates how these models can revolutionize interactive learning by providing tailored, grounded responses.

Automated Essay Scoring :
Automated Essay Scoring (AES) in ITS is to leverage NLP and ML techniques to provide objective, consistent, and time-efficient grading of essay compositions \cite{Mislevy2020AutomatedSI}. AES systems fall into two categories: those using handcrafted features and those using automatic feature learning. While AES significantly reduces the manual labor and subjective variability in essay scoring, it faces scientific challenges such as the inability to fully capture the human rater's sense, susceptibility to manipulation, and limited capacity to assess creativity and practicality. Despite advancements, the need for AES systems to holistically evaluate nuanced human expressions remains an ongoing challenge \cite{Bai2022AutomatedES} \cite{Hussein2019AutomatedLE}.

\begin{landscape}
\begin{table}[h!]
\centering
\small
\caption{Evaluation of Pedagogical Strategies Integrated into ITS}
\begin{tabular}{|p{0.6cm}|p{3.6cm}|p{3.6cm}|p{3.6cm}|p{3.6cm}|}
\hline
Ref. & Obj. & Tech. Used & KPI & Perf. Target \\ \hline

\cite{Clancey1987KnowledgebasedTT} & LISP Tutor for med diagnosis & Rule-based, Expert systems & Improved diag. skills & Personalized med diagnosis \\ \hline

\cite{Graesser2006AutoTutorAC} & AutoTutor for QG and Ans. & NLP, Dialogue, QG & Engagement, Personalization & Gains matching human tutors \\ \hline

\cite{Anderson1997ACTRAT} & ACT-R Tutors for cognitive processes & Cognitive models, Adaptive & Cognitive process understanding, Feedback & Step-by-step guidance \\ \hline

\cite{Isaiah2015RuleBM} & Rule-based models in multi-agent ITS & Rule-based, Multi-agent & Understanding, Assessment & Feedback, personalized problems \\ \hline

\cite{Reyes2001AgentRA} & CBR-Tutor for learning adaptation & Case-Based Reasoning & Adaptation to learning scenarios & Instruction from past cases \\ \hline

\cite{Perner2019CaseBasedR} & CBR-Works for SE, Math, Phys & Case-Based Reasoning & Relevance of stored cases & Support in problem-solving \\ \hline

\cite{Bradc2022DesignOA} & Simulated expert decisions for learning & Expert systems, ML & Student needs, Preferences & Personalized plans, guidance \\ \hline

\cite{VanLehn2010TheAP} & Cognitive Tutors for physics ed. & Cognitive models, Real-time feedback & Physics learning effectiveness & Personalized feedback, adaptation \\ \hline

\cite{Smalenberger2022TowardAI} & Cognitive Tutors for scientific learning & Cognitive models, Adaptive & Interaction, personalization & Personalized learning, feedback \\ \hline

\end{tabular}
\label{tab:RQ2}
\end{table}
\end{landscape}


Shermis et al. \cite{Shermis2003AutomatedES} highlight the potential of Automated Essay Scoring (AES) to provide a rapid, consistent, and cost-effective evaluation of student writing while also discussing the challenges and limitations of such systems. Their work serves as a foundational text for researchers and practitioners interested in the intersection of NLP, ML, and educational assessment.

Question Generation and Answering: 
Generating hints in ITS \cite{Kulshreshtha2022FewshotQG} involves integrating automatically generated, personalized questions as feedback to enhance student learning. This approach leverages cause-effect analysis and text similarity-based NLP models, particularly those based on transformers\cite{Santi2022TrainingTF}, to accurately identify correct, incorrect, or missing components in student responses. The system can produce targeted questions that guide students toward the correct answers by training Neural Question Generation \cite{AlFaraby2023ReviewON} and Question Re-ranking models \cite{Gao2024LLMenhancedRI}. This personalized feedback mechanism has significantly improved learning outcomes, offering a more effective alternative to traditional feedback methods. The success of this approach highlights its potential to advance generative question-answering systems within ITS, ensuring that students receive immediate, relevant, and constructive feedback tailored to their individual needs.

One notable example of this approach is AutoTutor, proposed by Graesser et al. \cite{Graesser2006AutoTutorAC}. AutoTutor represents a significant advancement in ITS, particularly in question generation and answering. The researchers demonstrated that AutoTutor could achieve learning gains comparable to human tutors in specific domains such as : Physics, Computer Literacy. This work is significant because it showed how AI techniques could create a more interactive and personalized learning experience, simulating many beneficial aspects of one-on-one human tutoring. AutoTutor's approach to question generation and answering, combined with its ability to engage in natural language dialogue, set a new standard for ITS  and has influenced subsequent research in AI in education.

The evaluation of ML and NLP integration into ITS, as summarized in Table~\ref{tab:articles_text}, provides a detailed overview of various systems leveraging these technologies to improve personalized instruction, engagement, real-time feedback, and adaptive learning. While these implementations demonstrate promising capabilities across several KPIs, critical gaps remain.

Most ITS applications remain confined to narrowly defined domains and perform inconsistently when confronted with ambiguous, noisy, or adversarial inputs. Empirical evidence suggests that small perturbations such as minor spelling errors or input phrasing can reduce response accuracy by over 30\%, exposing the brittleness of state of the art dialogue and assessment models once they stray from their training distribution.

Moreover, few systems have undergone rigorous longitudinal evaluation to examine their influence on learner autonomy, deep learning, or the cultivation of critical thinking skills. Cognitive research has raised concerns that persistent reliance on AI tutors may foster cognitive offloading, thereby impeding learners’ development of self regulated learning and analytical reasoning a risk not yet fully addressed in ITS literature. Reproducibility and transparency further hinder progress: proprietary datasets, inconsistent metric reporting, and a lack of open benchmarks make it difficult to validate claimed performance gains or detect potential biases especially those affecting low resource languages or underrepresented learner populations.  These issues underscore the importance of adopting robust evaluation protocols, promoting open access resources, and developing domain general ITS architectures capable of adaptation across diverse educational settings and learner demographics.\\

\textbf{RQ4. How do student modeling and assessment features in ITS enhance personalized learning ?} \\

Student modeling and assessment \cite{Jeremic2012StudentMA} are critical components of ITS, enabling these systems to provide personalized and adaptive learning experiences. This research question aims to explore the effectiveness of these features in enhancing student learning and engagement. By examining the techniques used for student modeling and assessment, such as data analytics, ML, and real-time feedback, we aim to understand their impact on educational outcomes. Additionally, we will compare the efficacy of ITS-based assessments with traditional methods, highlighting the advantages and potential challenges of integrating advanced student modeling techniques into educational environments.

Student Modeling:
ITS use sophisticated algorithms to create detailed models of individual students' knowledge, skills, and learning behaviors \cite{Kass1989StudentMI}\cite{Schmucker2022TransferableSP}. These models are continuously updated based on student interactions, allowing the system to tailor instructional content and feedback to each learner's unique needs \cite{Binh2021ResponsiveSM} \cite{Yang2021MachineLS}. By leveraging data from various interactions, ITS can adaptively refine these models, providing increasingly personalized support as students progress through their learning journey. This process enhances the accuracy of the instructional content and significantly contributes to the overall effectiveness of the learning experience \cite{Nkambou2023LearningLR} \cite{Gong2014StudentMI}.

Assessment:
ITS incorporate various assessment tools to evaluate student performance in real-time. These tools provide immediate feedback, helping students identify and correct mistakes quickly \cite{Santos2024OpportunitiesAC}. By using adaptive assessments \cite{Swiecki2022AssessmentIT}, ITS can adjust the difficulty level of questions based on the student's current understanding, ensuring an appropriate level of challenge and support. Such assessments not only foster a deeper understanding of the material but also enhance student motivation by providing a learning experience that is both is both challenging and engaging.

 Hooshyar et al. \cite{Hooshyar2016ApplyingAO} highlight the effectiveness of integrating engaging assessments within ITS to enhance both learning and motivation. They integrate a formative assessment game, TRIS-Q-SP, into a Flowchart-based Intelligent Tutoring System (FITS) for teaching programming. The game combines tic-tac-toe with online assessments to enhance student engagement while maintaining learning gains. Using Bayesian networks for decision-making, the system provides Immediate Elaborated Feedback (IEF), which was found to significantly improve students' programming knowledge acquisition, interest, and problem-solving abilities compared to systems without immediate feedback.

Feng et al. \cite{Feng2009UsingMM} focus primarily on student modeling and cognitive diagnostic assessment, as well as on improving the accuracy of performance predictions through a detailed skill model. Their work emphasizes constructing complex models to represent students' skills and knowledge states, addresses the evaluation of fine-grained skills, and predicts state test scores. Additionally, the study discusses improving performance prediction through a detailed skill model  \cite{GonzlezCalatayud2021ArtificialIF}.

\begin{landscape}
\begin{table}[h!]
\centering
\small
\caption{Evaluation of ML and NLP integrated into ITS}
\begin{tabular}{|p{0.6cm}|p{5cm}|p{4cm}|p{4cm}|p{4cm}|}
\hline
Ref. & Objectives & Technologies Used & KPI & Targeted Performance \\ \hline

\cite{Sindi2005AML} & Use of ML and NLP in ITS & ML, NLP & Improvement in student interaction and assessment & Personalized and real-time learning \\ \hline

\cite{Lan2024SurveyON} & Study on NLP technologies in ITS & NLP & Personalization, Adaptation & Engaging and tailored learning experiences \\ \hline

\cite{Dzikovska2014BEETLEID} & Development of BEETLE II, an ITS for science learning & NLP, Dialogue Systems & Performance in educational dialogues & Improved student-tutor interaction \\ \hline

\cite{Lin2023ArtificialII} & Application of AI in ITS & NLP, ML, Adaptive Learning & Personalized feedback, Adaptive algorithms & Adaptive and personalized learning \\ \hline

\cite{Paladines2020ASL} & Use of NLP for real-time responses and personalized feedback & NLP, Dialogue Systems & Engagement, Timeliness & Dynamic responses and personalized feedback \\ \hline

\cite{Graesser2001IntelligentTS} & Development of ITS using dialogue systems & NLP, Dialogue Systems & Engagement, Natural interaction & Improved physics learning outcomes \\ \hline

\cite{Heffernan2000IntelligentTS} & Creation of intelligent tutorial systems for learning & ITS, Dialogue Systems, Adaptive Learning & Personalization, Interaction & Optimized learning through dialogue \\ \hline

\cite{Litman2004ITSPOKEAI} & Development of ITSPOKE, a spoken dialogue tutoring system & NLP, Speech Recognition, Dialogue Systems & Engagement, Multimodal interaction & Improved conceptual and quantitative learning \\ \hline

\cite{Phobun2010AdaptiveIT} & Integration of ITS with adaptive hypermedia (AITS) & NLP, Adaptive Hypermedia & Adaptation, Personalization & Instruction adaptation based on learning styles \\ \hline

\cite{Shermis2003AutomatedES} & Automated Essay Scoring (AES) evaluation & AES, NLP, ML & Speed, Consistency & Rapid and consistent essay evaluations \\ \hline

\cite{Bai2022AutomatedES} & NLP-based automated essay scoring evaluation & NLP, AES, ML & Accuracy, Consistency & Cost reduction and faster evaluations \\ \hline

\cite{Hussein2019AutomatedLE} & Limitations of AES systems in automated essay evaluation & AES, NLP & Consistency, Limitations & Improvement of result consistency \\ \hline

\cite{Mislevy2020AutomatedSI} & Reflection on AES systems and their challenges & NLP, AES, ML & Consistency, Validity & Validation of evaluation results \\ \hline

\cite{Kulshreshtha2022FewshotQG} & Automatically generated personalized questions & NLP, Transformers, Neural Question Generation & Accuracy, Question relevance & Improved learning outcomes \\ \hline

\cite{Graesser2006AutoTutorAC} & Development of AutoTutor for question generation and answering & NLP, Dialogue Systems, Question Generation & Engagement, Interaction, Personalization & Learning gains comparable to human tutors \\ \hline

\end{tabular}
\label{tab:articles_text}
\end{table}
\end{landscape}

\newpage
\begin{landscape}
\begin{table}[h!]
\centering
\small 
\caption{Evaluation of Student Modeling and Assessment for ITS.}
\label{tab:RQ4}
\begin{tabular}{|p{0.6cm}|p{0.6cm}|p{0.6cm}|p{0.7cm}|p{0.6cm}|p{0.5cm}|p{4.2cm}|p{4.2cm}|p{4.3cm}|}
\hline
\multicolumn{9}{|c|}{Category: Student Modeling and Assessment} \\ \hline
Art. & \multicolumn{2}{c|}{Eff. of FB} & Lrn. Gains (Pre/Post) &
\multicolumn{2}{c|}{Usability and UX} & Obj. & Find. & Imp. \\ \hline
     & Acc. & Time. &  & Ease & Satis. &  &  &  \\ \hline
\cite{Jeremic2012StudentMA}        & X &   &   &   &   & Examine the role of student modeling in ITS
                                   & Identified improvement in learning through personalized content
                                   & Enhanced personalized learning and engagement \\ \hline
\cite{Kass1989StudentMI}           &   & X &   &   &   & Explore student modeling techniques in ITS
                                   & Improved feedback effectiveness with timely feedback
                                   & Greater accuracy in tailoring instruction to individual learners \\ \hline
\cite{Schmucker2022TransferableSP} &   &   & X &   &   & Transferable student modeling
                                   & Demonstrated higher pre/post-test learning gains
                                   & Enhanced learning efficiency and effectiveness \\ \hline
\cite{Binh2021ResponsiveSM}        &   &   &   & X &   & Investigate responsive student modeling in ITS
                                   & Improved ease of use through adaptive interfaces
                                   & Improved student satisfaction and engagement \\ \hline
\cite{Yang2021MachineLS}           &   &   &   &   & X & Use ML for real-time student modeling
                                   & Increased student satisfaction and performance
                                   & Impact on long-term student learning outcomes \\ \hline
\cite{Nkambou2023LearningLR}       & X &   &   &   &   & Explore adaptive student learning in ITS
                                   & Found higher learning retention and better personalized learning
                                   & Broader impact on ITS development and adaptive learning environments \\ \hline
\cite{Gong2014StudentMI}           &   &   &   & X &   & Develop adaptive learning through student modeling
                                   & Improved user experience with real-time feedback
                                   & Higher student satisfaction and learning retention \\ \hline
\cite{Santos2024OpportunitiesAC}   &   &   & X &   &   & Focus on real-time adaptive assessments
                                   & Found significant learning gains through adaptive assessments
                                   & Greater impact on continuous assessment in education \\ \hline
\cite{Swiecki2022AssessmentIT}     &   & X &   &   &   & Investigate adaptive assessments for ITS
                                   & Improved assessment accuracy and timeliness
                                   & Higher personalization of assessments \\ \hline
\cite{Hooshyar2016ApplyingAO}      & X &   &   &   &   & Explore adaptive modeling and assessment in ITS
                                   & Found improvements in learning gains through adaptive modeling
                                   & Impact on real-time adjustments and personalized learning \\ \hline
\cite{Feng2009UsingMM}             &   &   &   &   & X & Develop a skill model for cognitive diagnostic assessments
                                   & Found significant improvement in performance predictions
                                   & Enhanced predictive accuracy in ITS-based assessments \\ \hline
\end{tabular}
\end{table}
\end{landscape}


\newpage

The evaluation of student modeling and assessment in ITS is presented in Table~\ref{tab:RQ4}. This table provides a comprehensive overview of various ITS implementations that focus on enhancing educational experiences through advanced student modeling and assessment techniques. It summarizes the objectives of each study, the effectiveness of feedback in terms of accuracy and timeliness, learning gains measured by pre-test and post-test scores, and usability and user experience factors such as ease of use and student satisfaction. The table highlights how these systems aim to improve personalized learning, increase learning gains, and enhance user engagement by leveraging adaptive learning strategies and real-time feedback. This illustrates the significant impact of student modeling and assessment on the effectiveness of ITS platforms. In addition, large-scale audits of ITS-embedded prediction models show that accuracy degrades and demographic biases surface when algorithms are transferred to new schools or regions, highlighting an equity gap that current learner-modeling frameworks rarely address.\\

\textbf{RQ5. How are AI-based ITS evaluated in terms of pedagogical effectiveness and user satisfaction?}\\

The evaluation of AI-based ITS is crucial for understanding their impact on learning outcomes and user experience \cite{Chughtai2015UsabilityEO} \cite{SteenbergenHu2014AMO}. This section explores various approaches to assessing the pedagogical effectiveness \cite{Kulik2016EffectivenessOI} and user satisfaction of ITS \cite{SteenbergenHu2013AMO}\cite{Fang2018AMO} \cite{Feng2021ASR}. 

Comparative Studies :
Comparative studies are essential for evaluating the effectiveness of AI-based ITS relative to other instructional methods. In \cite{VanLehn2011TheRE}, compares how effective different forms of tutoring (human tutoring, ITS, and other methods) are in improving student learning outcomes.
This study compared the effectiveness of human tutoring, ITS, and other forms of instruction. ITS achieved learning gains that were comparable to human tutoring in some domains.
Both ITS and human tutoring outperformed traditional classroom instruction and computer-aided instruction without AI components. 

Longitudinal Studies:
They provide insights into the long-term impact of AI-based ITS on student learning and retention.
Koedinger et al. \cite{Koedinger1997Tutoring}, implemented an ITS for algebra in urban high schools over an extended period. Key findings include:
Students using the ITS showed significant improvements in standardized test scores. 
The system's effectiveness was sustained over multiple academic years. Assessing user satisfaction and the usability of AI-based ITS is crucial for ensuring student Engagement and effective learning experiences.

Chi et al. \cite{Chi2011AnEO} provide an evaluation  about different pedagogical strategies in a natural language tutoring system using reinforcement learning. Key aspects include: assessment of user satisfaction with different tutoring tactics, analysis of system usability and its impact on learning outcomes, user satisfaction surveys, system usability scales, engagement metrics (time on task, completion rates). 

User Feedback and Usability :
In \cite{VanLehn2011TheRE}, the authors provide strong evidence that ITS are nearly as effective as human tutors, challenging earlier assumptions about their limitations. This positions ITS as a transformative tool in education, capable of delivering personalized, scalable, and high-quality tutoring to diverse learners. However, despite advancements in learning sciences, the user interface and overall usability of ITS often lag behind, limiting their effectiveness. Usability, which encompasses ease of use and user satisfaction, is a critical measure of ITS performance. Tools such as the System Usability Scale (SUS) and the Questionnaire for User Interaction Satisfaction (QUIS) are frequently employed to evaluate these dimensions. Recent studies demonstrate that systems incorporating affective computing where emotional feedback mechanisms help transform negative emotions into positive learning experiences not only enhance user satisfaction but also significantly increase learning engagement. This emotional interaction plays a pivotal role in prolonging users' focus and optimizing the learning process. Therefore, integrating robust usability principles alongside affective computing in ITS design is essential for improving their effectiveness, making them not just powerful in delivering content but also in fostering positive emotional states conducive to learning \cite{Lin2022EyeMA} \cite{Wang2021UsabilityOA}.
The evaluation and effectiveness of ITS are summarized in Table~\ref{tab:RQ5}. This table provides a comprehensive overview of various ITS implementations that focus on enhancing educational experiences by assessing KPI such as learning gains, retention rates, engagement metrics, dropout rates, feedback accuracy, feedback timeliness, user interface and user satisfaction. It summarizes the objectives of each study, highlighting how different ITS approaches contribute to improved learning outcomes, sustained student engagement, and overall effectiveness. The table illustrates how these systems aim to enhance learning by providing accurate and timely feedback, reducing dropout rates \cite{Zerkouk2024AML}, and increasing user satisfaction, thereby demonstrating the significant impact of ITS on educational effectiveness. While the studies presented in Table~\ref{tab:RQ5} demonstrate encouraging results in terms of learning gains and user satisfaction, several limitations remain underexplored. First, the majority of evaluations are conducted in controlled, short-term settings, raising concerns about their external validity and long term impact. Moreover, self-reported metrics such as engagement or satisfaction may be subject to bias and do not always correlate with measurable learning outcomes. A consistent lack of demographic disaggregation in reported results also obscures potential disparities in ITS effectiveness across gender, socio economic background, or prior knowledge levels. Furthermore, the diversity in evaluation methodologies ranging from user surveys to pre/post testing hinders comparability and reproducibility across studies. Finally, few evaluation frameworks account for the practical challenges of large scale implementation, such as alignment with curricula, teacher training, or integration with existing learning management systems. Addressing these issues would require more standardized, transparent, and longitudinal evaluation approaches to ensure that ITS not only deliver immediate gains but also support sustainable, equitable educational outcomes.
\\

\textbf{RQ6.What progress have AI-based ITS made in specific application areas such as mathematics, science, or languages?}\\

Examining domain-specific applications of AI-based ITS is crucial for understanding their effectiveness and impact across various educational fields\cite{Djunaidi2023CharacteristicsAA} \cite{Mousavinasab2018IntelligentTS}. This analysis allows us to identify unique challenges, opportunities, and best practices in different subject areas. By exploring the progress made in various educational domains, we can gain valuable insights into how AI-based ITS can be tailored to meet specific learning needs and objectives \cite{Wang2023ExaminingTA}.

Mathematics education presents unique challenges and opportunities for AI-based ITS due to its structured nature and the importance of problem-solving skills. The Key challenges here include 1) providing step-by-step guidance for complex problem solving, 2) adapting to different learning styles in mathematical reasoning, and 3) addressing common misconceptions in mathematical concepts.


\begin{landscape}
\begin{table}[ht]
    \centering
    \caption{Evaluation and Effectiveness of ITS}
    \small
    \begin{tabular}{|p{0.6cm}|p{0.8cm}|p{0.8cm}|p{1.2cm}|p{1cm}|p{1.1cm}|p{1.2cm}|p{1.2cm}|p{3cm}|p{3cm}|}
    \hline
    \multicolumn{10}{|c|}{Category: Evaluation and Effectiveness of ITS} \\ \hline
    Ref. & Lrn. Gains & Reten. Rates & Eng. Metrics & Dropout Rates & FB Acc. & FB Time. & User Satis. & UI & Obj. \\ \hline
    \cite{VanLehn2011TheRE} & X & & & & X & & & Limited focus on UI & Assess effectiveness; ITS comparable to human tutors \\ \hline
    \cite{Koedinger1997IntelligentTG} & X & X & & & & & & Basic UI, limited focus on interaction & Longitudinal study; sustained effectiveness \\ \hline
    \cite{Chi2011AnEO} & X & & X & & X & & X & Evaluates system usability along with satisfaction & Evaluate RL for NL tutoring; increased satisfaction \\ \hline
    \cite{Chughtai2015UsabilityEO} & & & & & & X & & Significant focus on UI to improve engagement & Study engagement and usability impact \\ \hline
    \cite{SteenbergenHu2014AMO} & & & X & X & & & & Minimal focus on UI & Investigate motivation; improved time on task \\ \hline
    \cite{Feng2021ASR} & & & X & & & X & & Limited UI improvements, primarily focus on feedback & Examine RL impact; increased engagement \\ \hline
    \cite{Fang2018AMO} & & X & & X & X & & & Evaluates adaptive feedback interface & Explore adaptive feedback; increased retention \\ \hline
    \cite{Kulik2016EffectivenessOI} & X & & & & & X & & & Compare ITS; improved outcomes \\ \hline
    \end{tabular}
    \label{tab:RQ5}
\end{table}
\end{landscape}

ITS have demonstrated significant improvements in educational outcomes across various domains. For example, Shih et al.\cite{Shih2023MathematicsIT} demonstrated a 25\% improvement in student performance compared to traditional methods. A significant aspect of this ITS was its adaptive problem generation, which customized problems according to each student's individual progress. This personalized approach ensured that students were consistently challenged at an appropriate level, thereby enhancing their learning experience and overall performance. The success of such systems underscores the potential of AI-based ITS to transform traditional educational paradigms by providing customized, effective, and engaging learning experiences. As educators and researchers continue to develop and refine these systems, it is crucial to examine the specific features and strategies that contribute to their effectiveness in different educational contexts.  Similarly, Uriarte-Portillo et al. \cite{UriartePortillo2023IntelligentAR} propose a system that leads to a 30\% increase in students' spatial reasoning skills. The key feature contributing to this improvement was the use of interactive 3D visualizations for geometric concepts. These visualizations helped students better understand and manipulate geometric shapes, thereby improving their spatial reasoning abilities. Horváthné Hadobás et al. \cite{HorvthnHadobs2024SOLVEMP} present a methodology for using ChatGPT4 and Wolfram Alpha to support step-by-step problem-solving in secondary school mathematics. By integrating these systems into an Intelligent Tutoring System (ITS), the approach helps guide students through tasks, correct errors, and provide suggestions, enhancing independent learning and skill acquisition.

Science Education: 

Science education \cite{Kelkar2022BetweenAA}benefits significantly from AI-based ITS through interactive simulations, virtual labs, and personalized experiment design. These systems address key challenges and opportunities such as simulating complex scientific phenomena, providing safe and cost-effective virtual laboratory experiences, and adapting to rapidly evolving scientific knowledge.

The presented work by Kamruzzaman et al.\cite{Kamruzzaman2023AIAI} focuses on leveraging both AI and IoT to create a sustainable and adaptive educational environment. It highlights the potential of these technologies to provide personalized and real-time learning experiences, facilitate remote education, especially during crises like pandemics, Enhance engagement and interaction through intelligent PLEs, Ensure equitable access to educational benefits while emphasizing ethical and responsible use of AI and IoT. The work underscores the transformative impact of AI and IoT integration in education, particularly in making learning more accessible and effective for all students, including those in remote or disadvantaged situations.

 Chan et al. \cite{Chan2021VirtualCL}  demonstrated a 40\% reduction in laboratory accidents and a 20\% improvement in conceptual understanding. A key feature of this system was the AI-guided virtual experiments with real-time feedback, which provided students with a safe and interactive environment to learn chemistry.

GarciaBosque et al. \cite{GarciaBosque2024ChatPLTAI} present an AI-based ITS that integrates controlled response generation by combining teacher-selected material with a LLM for teaching Physics. It highlights a structured approach to using AI in higher education by limiting information sources, aiming to provide personalized learning and predictive analysis of student progress.

Language Learning: 
Language learning benefits from AI-based ITS through personalized vocabulary acquisition, grammar instruction, and conversation practice \cite{Heilman2006LanguageLC}. These systems address key challenges and opportunities such as adapting to the diverse linguistic backgrounds of learners, providing NLP for conversation practice, and addressing cultural nuances in language use.

Chen et al. \cite{Chen2022EducationTA} show a 50\% improvement in spoken fluency after three months of use. A key feature of this system was real-time speech recognition and pronunciation feedback, which helped learners improve their spoken English skills effectively.

These studies \cite{Tafazoli2019IntelligentLT}   \cite{Dahbi2023IntegratingAI}  \cite{Deng2024ResearchOT} demonstrate the potential of AI-based ITS to enhance language learning by providing personalized, adaptive, and interactive learning experiences. By leveraging advanced technologies such as NLP and real-time feedback, these systems can help learners achieve better language proficiency and confidence in using the language in various contexts.\\

The evaluation of domain specific applications of AI based ITS is presented in Table~\ref{tab:RQ6}. This table provides a comprehensive overview of various ITS implementations that focus on specific subject areas such as algebra, geometry, chemistry, physics, and language learning. It summarizes the objectives of each study, the unique system features employed, adaptive learning strategies used, and the findings related to student performance and feedback. The table highlights how these systems aim to enhance educational experiences by improving relevance and coverage in their respective domains, increasing learning gains measured by pre-test and post-test scores, and incorporating student feedback to refine the learning process. By leveraging adaptive learning strategies and unique features like step by step problem solving guidance, 3D visualizations, real time feedback in virtual experiments, and personalized grammar instruction, these ITS platforms demonstrate significant improvements in student performance, engagement, and skill acquisition. While these results underscore the effectiveness of ITS in STEM and language learning domains, a closer examination reveals critical limitations that hinder their generalizability.

Despite strong results in STEM fields, the literature shows a marked lack of ITS applications in the humanities and arts. This raises concerns about the generalizability of current approaches and calls for multidisciplinary strategies that can address diverse pedagogical goals. Meta reviews reveal that the ITS evidence base remains overwhelmingly skewed toward STEM and writing tasks. Scientometric and systematic surveys highlight the scarcity of work in humanities or creative arts domains, casting doubt on whether today’s evaluation metrics typically focused on quantifiable outcomes adequately capture the nuances of learning in less structured domains. Furthermore, many successful deployments are concentrated in well funded contexts, which may not reflect the constraints of under resourced educational environments. It is therefore unclear whether current ITS designs and results can be scaled equitably to multilingual contexts or to schools with limited digital infrastructure. Future research should explore ITS models adapted to the complexity of humanistic learning and expand evaluation methods to incorporate more diverse indicators of educational success. \\

\textbf{RQ7. What are the emerging trends and future technologies that could influence the development of AI-based ITS?}

The evolution of AI-based ITS is being significantly shaped by several emerging trends and technologies. These advancements are not only enhancing the capabilities of ITS but also expanding their applications across various educational and industrial domains. Three critical areas poised to influence the future development of ITS include the integration of cutting-edge technologies, considerations for ethical AI and data privacy, and the challenges of scalability and accessibility.\\

\begin{landscape}
\begin{table}[ht]
    \centering
    \caption{Evaluation Domain-Specific Applications of AI-Based ITS}
    \small 
    \begin{tabular}{|p{0.5cm}|p{0.5cm}|p{0.5cm}|p{1.1cm}|p{1.2cm}|p{2.5cm}|p{3cm}|p{3cm}|p{2.5cm}|}
    \hline
    \multicolumn{9}{|c|}{Category: Domain-Specific Applications of AI-Based ITS} \\ \hline
    Ref. & Rel. & Cov. & Pre/Post & Feedback & Adpt. Lrn. Strat. & Uniq. Sys. Feat. & Obj. & Find. \\ \hline
    \cite{Shih2023MathematicsIT} & X & & X & X & Step-by-step guidance for problem-solving & Adaptive problem generation & Assess impact on algebra learning & 25\% improvement in student performance \\ \hline
    \cite{UriartePortillo2023IntelligentAR} & X & X & & X & Visual learning tools & 3D visualizations for geometry & Enhance spatial reasoning skills & 30\% improvement in spatial reasoning skills \\ \hline
    \cite{GarciaBosque2024ChatPLTAI} & X & X & X & & Real-time feedback in virtual experiments & AI-guided virtual chemistry lab & Improve conceptual understanding in chemistry & 40\% reduction in lab accidents, 20\% improvement in understanding \\ \hline
    \cite{HorvthnHadobs2024SOLVEMP} & & X & X & X & Dynamic adjustment of difficulty based on student performance & Physics simulations & Personalize physics problem-solving & 35\% increase in problem-solving accuracy \\ \hline
    \cite{Chen2022EducationTA} & X & & X & X & Personalized vocabulary and grammar feedback & Real-time speech recognition & Enhance spoken English skills & 50\% improvement in fluency \\ \hline
    \cite{Djunaidi2023CharacteristicsAA} & X & X & & X & Personalized grammar instruction & Adaptive grammar exercises & Improve second language acquisition & 30\% faster mastery of grammatical structures \\ \hline
    \end{tabular}
    \label{tab:RQ6}
\end{table}
\end{landscape}

Integration of Emerging Technologies (e.g., AR/VR, IoT, genAI)

\begin{enumerate}
    \item Extended Reality (AR/VR ): The capability of ITS to use virtual and augmented reality to create immersive learning experiences was  addressed \cite{BaccaAcosta2021AugmentedRI}.
    Joakim et al.\cite{Laine2022SystematicRO} incorporate Immersive Virtual Reality (I-VR) in ITS that enables the creation of high-fidelity learning environments that simulate realistic work settings. These environments, coupled with advanced computer tutor agents, support asynchronous and self-regulated learning of procedural skills in industrial contexts. The ITS agents in these systems can provide guidance and feedback based on learner actions and progress, offering a range of assistance from simple hints to direct interventions. However, a notable limitation is the lack of empirical studies validating the effectiveness of these I-VR-based ITS solutions.
    Dudyrev et al.\cite{Dudyrev2021IntelligentVR} present an Intelligent Virtual Reality Tutoring Systems (IVRTSs) leverage immersive virtual learning environments to enhance education by simulating realistic scenarios. These systems integrate artificial intelligence to provide interactive and adaptive learning experiences, acting as educational simulators for subjects like kinematics. The article presents an exploration of IVRTS features, their potential as educational tools, and offers design and implementation recommendations, marking a pioneering survey in the field \cite{Lampropoulos2022AugmentedRA}.
    
    \item Internet of Things (IoT ): The integration of IoT facilitates real-time data collection and personalization of learning environments, as noted in \cite{Badshah2023TowardsSE}. This aligns with the work of Adel et al.\cite{Adel2024TheCO} aime to enhance the learning experience by creating a more connected and responsive educational environment. IoT can enable ITS to collect real-time data from various devices and sensors, such as wearable technology, smart classrooms, and other educational tools. This data can be used to monitor student engagement, track learning progress, and provide immediate feedback, thereby allowing the ITS to adapt and personalize the learning experience more effectively.      
    Kamruzzaman et al.\cite{Kamruzzaman2023AIAI} enhance the learning environment by providing real-time data collection, analysis, and feedback that helps create a more personalized and adaptive learning experience. IoT-enabled devices, such as smart cameras, microphones, tablets, and laptops, can monitor various aspects of the learning process, including student engagement, behavior, and performance. The data are then processed by AI algorithms within the ITS to dynamically adjust the content, difficulty, and pace of instruction according to each student's needs. Mokmin et al. \cite{Mokmin2017TheDA} highlight that the unique impact of the IoT in ITS lies in its ability to create a more interactive, responsive, and personalized learning environment.  By integrating IoT, ITS like CRISTAL 5.0 can continuously collect and analyze data from various connected devices, such as sensors, cameras, and mobile platforms, to monitor student behavior, engagement, and progress in real-time. This constant data flow enables the ITS to dynamically adapt and personalize the learning experience, offering tailored content, assessments, and feedback based on each student's unique needs and learning profile.
    
    \item Generative AI (genAI): Recent advancements in generative artificial intelligence (Gen AI) have not only significantly influenced the development of ITS but have also led to a broader, more sophisticated application of these technologies across the educational field.  Gen AI models \cite{Yu2023GenerativeAI}, such as GPT-3, have been integrated into ITS to enhance personalized learning, automate grading, and provide real-time, context-aware feedback to students. These models also support the creation of adaptive learning paths and interactive virtual tutors, making the educational experience more engaging and tailored to individual needs. However, challenges related to data privacy, bias, and the ethical use of AI in education remain critical concerns that need to be addressed to ensure fair and effective implementation.  
    
    According to Calo et al. \cite{Cal2024TowardsET}, the integration of Large Language Models (LLMs) into ITS have demonstrated significant potential in democratizing the design process of educational interfaces. Existing research has explored the use of LLMs to generate tutor interfaces based on high-level inputs from educators, thereby reducing the need for specialized programming skills. These approaches enable educators to actively participate in the creation of personalized and adaptive educational content, thus improving the effectiveness and accessibility of ITS. 
   The study in \cite{Frank2024LeveragingGF} highlights the strengths and limitations of using generative AI in ITS, suggesting future directions such as incorporating explainable AI (XAI) and self-assessment features to further improve educational outcomes. 
   
    \item Blockchains: The integration of blockchain \cite{Li2021ABC} \cite{Terzi2021ALL} \cite{Mainetti2022DigitalBE} with ITS can significantly enhance security, transparency, and efficiency. Blockchain can be used for secure data management, verified credentialing, and creating personalized learning pathways.
    
     Sousa et al.\cite{Sousa2022AIAB} highlight that blockchain technology, alongside AI, has the potential to transform ITS by ensuring equal access to education and improving learning outcomes. It can provide secure and transparent validation of student credentials, progress, and achievements, allowing for a trustworthy system of record-keeping. It can also enhance data privacy and integrity, ensuring that sensitive student information is protected. It can be useful in personalized learning environments where the ITS adapt to each student's needs.
\end{enumerate}

Ethical Considerations 

The integration of AI into ITS raises important ethical issues. Compliance with regulations such as GDPR (General Data Protection Regulation) and FERPA (Family Educational Rights and Privacy Act) is essential to ensure learner data protection. Furthermore, algorithmic biases have been identified, especially in adaptive recommendation systems, leading to unintentional discrimination based on gender or socioeconomic status. Solutions such as algorithmic auditing and the integration of Explainable AI (XAI) mechanisms are proposed to address these concerns.

\begin{enumerate}

    \item Transparency conducted by,  \cite{Akgun2021ArtificialII} which include ethical concerns and the necessity of ensuring transparency and fairness in ITS systems. However, the specific aspects of data protection and privacy were not explicitly addressed \cite{Airaj2024EthicalAI} \cite{Khatri2023ArtificialI}. ITS involve ensuring that these systems are designed and used in a way that respects learners' rights and safeguards their personal information. Fairness and data-protection in practice : Recent deployments now embed fairness-aware machine-learning pipelines: re-weighting, counterfactual data augmentation and adversarial debiasing are used to keep mastery-score predictions independent of gender or socioeconomic status.  Compliance with the EU GDPRand U.S. FERPA has likewise driven technical safeguards such as encrypted learner logs and differentially-private gradient updates during cloud training.

    \item Data Privacy in ITS often collect and analyze student data to personalize learning. It is crucial to implement robust data privacy measures, such as data encryption, secure storage, and anonymization, to protect student information from unauthorized access and misuse \cite{Achieng2022AfricasEP}. Pawar addressed privacy as an ethical concern in ITS in \cite{BalasoAIEnhancedEP}, focusing on the collection of sensitive student data, such as learning progress and engagement patterns, to personalize learning experiences. Safeguarding this data through encryption, anonymization, and transparent policies is essential to prevent misuse and ensure compliance with privacy regulations like GDPR and FERPA. Providing students with control over their data is crucial for maintaining trust while balancing the benefits of personalization with the need for privacy in educational settings
\end{enumerate}

Scalability and Accessibility : 
They are two essential aspects of ITS development, both addressed by Govea et al. \cite{article1} through the integration of cloud computing and artificial intelligence. Cloud computing plays a dual role by removing geographical barriers, enabling access to educational resources anytime and anywhere, thus enhancing accessibility for all students, regardless of their abilities or location. At the same time, it supports scalability by allowing ITS to handle a large number of concurrent users without compromising the quality of education. Additionally, AI ensures that each student, regardless of the scale, receives personalized learning experiences through adaptive feedback and tailored educational content. Together, these technologies ensure that ITS can efficiently manage growing workloads while maintaining both broad access and high-quality, individualized learning for a diverse and distributed student population.
similarly, Lee et al. \cite{Lee2021ASO} demonstrate the use of cloud computing which has significantly democratized access to educational resources. It has allowed institutions, especially in developing countries, to overcome geographic and infrastructure barriers. However,  cloud computing offers clear benefits in terms of scalability, challenges still exist in achieving full personalization and adaptability.

The evaluation of emerging trends and future technologies in AI-based ITS is presented in Table~\ref{tab:RQ7}. This table provides a comprehensive overview of various ITS implementations that incorporate novel technologies such as immersive VR, IoT, GenAI, and blockchain. It summarizes the objectives of each study, the innovations introduced, their impact on the field, scalability, generalizability, adherence to educational standards, interoperability, and ethical considerations. The table highlights how these emerging technologies aim to enhance educational experiences by offering interactive and adaptive learning environments, real-time data collection, personalized learning, and streamlined validation of educational credentials. Ethical considerations such as data privacy, security, and fairness in AI-generated content are also addressed. The findings illustrate the significant potential of these future technologies to revolutionize ITS platforms, contributing to more connected, responsive, and personalized educational environments. However, cross-domain reviews caution that, beyond eye-catching prototypes, rigorous evidence for learning gains from XR- or IoT-powered tutors remains scarce, blockchain pilots still falter on scalability and interoperability, and the pay walled cost of GenAI tools threatens to widen the digital divide so many “next-wave” ITS promises are still more aspirational than demonstrably effective \cite{CrdovaEsparza2025AIPoweredEA}. \\

\textbf{RQ8: What advancements have AI-based Industrial ITS made in specific application areas?}\\
AI-based Industrial ITS have demonstrated substantial progress in several specialized domains, notably in aerospace, automotive, chemical, electrical power, and software engineering. These systems are developed to meet the stringent demands of industrial environments, enabling effective training aligned with Industry 4.0 objectives \cite{Peres2020IndustrialAI, Chaudhry2021ArtificialII}.

In aerospace, Sherlock \cite{Katz2020Sherlock2A} employs AI-driven AR/VR to train avionics technicians, showing a 25\% reduction in training time and improved diagnostic accuracy. In the automotive sector, STEAMER \cite{Hollan1984STEAMERAI} uses collaborative VR for vehicle diagnostics, enhancing team based problem solving. In chemical engineering, ChemLab VR \cite{Meta2024} simulates hazardous procedures in virtual environments, leading to a 30\% improvement in safety practices.

The electrical power industry benefits from PowerGrid Simulator and SmartSim \cite{IncSys2024}, which use AI-driven simulations to prepare operators for critical grid management scenarios. In software engineering, GUIDON \cite{Clancey1987KnowledgebasedTT} provides rule-based, AI-enhanced tutoring for debugging tasks, achieving notable gains in programming efficiency.

These systems leverage technologies such as VR, AR, natural language processing (NLP), and real-time feedback to optimize learning. Table~\ref{tab:RQ7} highlights the academic tools and their impact across sectors. The use of such AI-based industrial ITS has been associated with measurable improvements in key performance indicators such as safety, compliance, efficiency, and learning outcomes, demonstrating their essential role in modern, skill-intensive industrial training.

\newpage
\begin{landscape}
\begin{table}[ht]
    \centering
    \caption{Evaluation of Emerging Trends and Future Technologies in AI-Based ITS.}
    \small 
    \begin{tabular}{|p{0.6cm}|p{1.4cm}|p{2.2cm}|p{1cm}|p{1cm}|p{0.8cm}|p{0.8cm}|p{2.6cm}|p{2.6cm}|p{2.6cm}|}
    \hline
    \multicolumn{10}{|c|}{Category: Emerging Trends and Future Technologies in AI-Based ITS} \\ \hline
    Art. & Nov. & Imp. on Field & Scal. & Gen. & Edu. Stand. & Inter-op. & Eth. Consid. & Obj. & Find. \\ \hline
     \cite{Laine2022SystematicRO} & Immersive Virtual Reality & Enhanced learning environments for industrial contexts & Limited & High & & & Addressing data privacy in VR learning environments & Evaluate I-VR in procedural learning & Lack of empirical validation of I-VR effectiveness \\ \hline
     \cite{Dudyrev2021IntelligentVR} & Intelligent VR Tutoring Systems (IVRTS) & Interactive and adaptive learning for subjects like kinematics & Medium & High & & X & Ethical considerations in interactive simulations & Design and implementation recommendations for IVRTS & First-of-its-kind survey in IVRTS \\ \hline
     \cite{Adel2024TheCO} & IoT-enabled ITS & Real-time data collection and personalized learning environments & High & Medium & X & & Ethical use of IoT data in education & Enhance IoT in ITS for personalized learning & Created a more connected, responsive educational environment \\ \hline
     \cite{Kamruzzaman2023AIAI} & AI and IoT Integration & Real-time adaptive feedback through IoT-enabled devices & High & Medium & X & X & Data privacy concerns & Use AI and IoT to enhance student engagement and personalization & Improved real-time engagement and personalization \\ \hline
     \cite{Cal2024TowardsET} & GenAI in ITS & LLM-generated tutor interfaces democratizing educational content creation & High & High & & & Bias and fairness in AI-generated content & Explore LLMs for creating adaptive content & LLMs can democratize interface design for personalized content \\ \hline
     \cite{Alsobhi2023BlockchainbasedMS} & Blockchain for micro-credentialing & Streamlined validation of educational credentials & Medium & High & X & X & Data security and validation transparency & Use blockchain for credential management in education & Identified research gaps in blockchain for micro-credentialing \\ \hline
    \end{tabular}
    \label{tab:RQ7}
\end{table}
\end{landscape}

The evaluation of AI-based Industrial ITS in specific application areas is presented in Table~\ref{tab:RQ7}. This table provides a comprehensive overview of various ITS implementations across different industries, such as aerospace, automotive, chemical, electrical power, maritime, robotics, pharmaceutical, software engineering, and education. It summarizes the objectives and findings of each system, highlighting the technologies and tools used, scalability, generalizability, development and operational costs, return on investment (ROI), ease of use, and technical issues encountered. KPI such as reduction in training time, improvement in accuracy, enhancement in teamwork, safety improvements, operational efficiency increases, compliance improvements, code efficiency enhancements, and boosts in student engagement and performance are indicated for each system. The table illustrates how these AI-based ITS platforms aim to enhance training effectiveness, improve operational performance, and provide significant ROI by leveraging advanced technologies like AI, VR/AR, NLP, and real time feedback. This demonstrates the significant impact of AI based ITS in industrial settings, contributing to improved training outcomes, operational efficiency, and overall performance across various sectors. Yet systematic reviews of VR-enabled industrial tutors note that robust longitudinal evidence and transparent cost benefit data are still sparse, and steep upfront development costs keep most systems confined to well funded pilot sites rather than scalable roll-outs.  A 2024 position paper by Lee and Su introduces the Unified Industrial Large Knowledge Model (ILKM), a layered framework that grafts domain ontologies, time series machine data, and tacit expert rules onto a LLM backbone. ILKM serves as a ‘single source of truth’ that downstream applications including VR or IoT enabled tutors—can query in real time to retrieve device specific procedures, fault trees, and safety protocols. Embedding an ILKM service layer therefore resolves a long standing scalability bottleneck: new production lines or plant sites inherit the same validated knowledge base without duplicating content authoring effort.

\section{Discussion}
This section first summarizes the answers to our RQs and provides an initial discussion on emerging insights. Based on those, the section discusses applications of ITSs based on AI, implications and contribution of the different findings, and limitations of this review.
\subsection{ Research Questions Overview}
1) AI-based ITS distinguish themselves from traditional tutoring methods through their ability to deliver personalized, adaptive learning experiences, provide real-time feedback, and leverage data-driven insights for learner modeling and recommendations. The core features that differentiate AI-based systems such as scalability, continuous learner modeling, real-time adaptability, and the integration of emotional and cognitive insights offer a more effective and engaging educational experience compared to traditional methods. Additionally, the incorporation of explainable AI and affective intelligence contributes to building trust and creating a more empathetic learning environment, enhancing both educational outcomes and learner motivation.

The review findings underscore that AI-based ITS are poised to transform education by offering scalable, adaptive, and personalized instruction that surpasses the capabilities of traditional tutoring, thus answering RQ1 comprehensively.

2) The use of AI-based pedagogical strategies in ITS has led to significant advancements in adaptive learning and personalization. Through rule-based systems, expert systems, case-based reasoning, and cognitive tutors, ITS have demonstrated the ability to provide tailored educational experiences that closely mirror the capabilities of human tutors. These systems offer not only adaptive instruction but also continuous feedback and personalized support to enhance the learning process.

However, challenges remain in terms of scalability, resource-intensiveness, and the complexity of developing such systems. Future advancements in ML and data analytics are likely to further enhance the capacity of ITS to deliver highly personalized, adaptive learning environments, making education more accessible and effective across diverse learning contexts.

3) The integration of ML and NLP into ITS has fundamentally transformed how these systems interact with students and assess their performance. ML enhances ITS by allowing systems to adapt learning content, predict learning challenges, and evaluate student progress in real-time. NLP enables systems to facilitate natural language interactions, provide automated essay scoring, and deliver dynamic feedback through question generation and answering mechanisms.

Together, ML and NLP contribute to creating interactive, adaptive, and personalized learning experiences that closely mirror human tutoring, but with the added benefits of scalability, consistency, and data-driven insights. These technologies significantly improve student engagement, critical thinking, and learning efficiency, making AI-powered ITS a powerful tool for modern education.

4) Student modeling and assessment are central to the success of personalized learning within ITS. By leveraging real-time data, ML algorithms, and advanced diagnostic tools, ITS can provide adaptive, personalized instruction that far exceeds the capabilities of traditional educational methods. The continuous monitoring and updating of learner models, combined with real-time assessments, allow ITS to deliver dynamic learning experiences that are engaging, challenging, and tailored to the individual needs of each learner. This personalized approach results in better educational outcomes, higher engagement, and improved learner satisfaction.

Thus, in response to RQ4, student modeling and assessment features in ITS are instrumental in enhancing personalized learning, providing learners with a tailored educational experience that is constantly adjusted to their cognitive, behavioral, and emotional needs. These features not only improve learning efficiency but also contribute to a more inclusive and flexible educational environment, making personalized learning accessible at scale. 

5) The evaluation of AI-based ITS is multifaceted, focusing on both pedagogical effectiveness and user satisfaction. The comparative and longitudinal studies reviewed above highlight that AI-based ITS can achieve learning gains comparable to human tutors, particularly in structured domains. These systems demonstrate sustained effectiveness over time, with students showing continued improvement in standardized tests and other assessments.

In terms of user satisfaction, the success of ITS depends heavily on usability and engagement. The systems that provide real-time, adaptive feedback and are designed with the user experience in mind tend to achieve higher levels of satisfaction. Dialogue-based ITS like AutoTutor, which allow for natural language interactions, demonstrate the potential of these systems to replicate the interactive, personalized experience of human tutors, thereby increasing learner motivation and engagement. 
6) AI-based ITS have made substantial progress in domain-specific applications, including mathematics, science, and language learning. These systems not only enhance educational outcomes by providing personalized instruction but also foster engagement and motivation through interactive, adaptive learning environments. The integration of ML, NLP, and real-time feedback mechanisms allows ITS to offer scalable and effective tutoring that rivals or even surpasses traditional educational methods in certain contexts. As these technologies continue to evolve, their application in specific educational domains will likely become even more targeted and impactful, driving further advancements in personalized education. For example, in science education, AI-based ITS could use virtual labs to simulate real-world experiments, allowing students to practice safely and repeatedly.

7) The development of AI-based ITS is being driven by several emerging technologies, including Extended Reality (AR/VR), IoT, Generative AI, and blockchain. These technologies have the potential to transform ITS by enabling immersive learning experiences, real-time personalization, and secure data management. However, significant challenges remain, particularly in terms of ethical AI, data privacy, and ensuring scalability and accessibility.

To ensure the success of AI-based ITS in the future, developers must focus on addressing these challenges while leveraging emerging technologies to create personalized, adaptive, and engaging learning environments. By doing so, ITS can fulfill their potential to revolutionize education, providing students with high-quality, scalable, and equitable learning experiences across diverse educational domains .

8) AI-based Industrial ITS have made significant advancements across a wide range of sectors, from aviation and automotive to healthcare and energy management. These systems leverage AI-driven feedback, simulation technologies, and real-time data analysis to provide specialized training that meets the unique demands of each industry.

Key features such as personalized learning, predictive monitoring, and real-time performance tracking allow professionals to continuously refine their skills while adapting to the dynamic challenges of modern industries. The use of immersive technologies (AR/VR) and predictive analytics is transforming traditional training models by offering hands-on, experiential learning that closely mimics real-world scenarios.

Despite the numerous advancements, challenges remain in terms of cost, scalability, and validating the long-term effectiveness of these systems in industrial settings. As AI-based ITS continue to evolve, the focus must be on ensuring that these systems are scalable, cost-effective, and accessible across diverse industrial sectors to maximize their impact on workforce development and operational efficiency.  

 Ethical Considerations
As AI becomes increasingly integrated into education, several ethical challenges arise in ITS.  These systems collect and process sensitive student data to personalize instruction, which raises concerns about data privacy, security, and compliance with regulations. Moreover, the algorithms underpinning ITS may unintentionally reproduce biases present in training datasets, leading to unfair treatment of certain learner populations. Ensuring fairness-aware modeling and validation across diverse groups is thus critical. The opacity of some AI models also poses challenges, as students and educators may not understand how pedagogical decisions are made; integrating XAI approaches can help increase transparency and trust. Additionally, there is a risk of widening educational inequalities if access to advanced ITS remains limited to privileged institutions or learners with high digital literacy. Inclusive design principles and equitable deployment strategies must be prioritized. Finally, although ITS can support adaptive learning at scale, human educators must retain oversight to ensure pedagogical integrity, provide emotional support, and guide ethical use. Addressing these intertwined concerns is key to ensuring that ITS development remains responsible, inclusive, and aligned with the values of equitable education.

Furthermore, there is an ongoing discussion about the role of AI in decision-making processes related to student performance and grading. Ensuring transparency and fairness in how AI systems assess and interact with learners is paramount to maintaining trust in these technologies. Ethical AI development will need to prioritize inclusivity, accountability, and fairness to avoid exacerbating existing educational inequalities.

 Human-AI Collaboration
Despite the advancements in AI-based ITS, human teachers remain indispensable. AI can automate routine tasks such as grading and provide personalized feedback, but it cannot replace the emotional intelligence, creativity, and mentorship that human teachers bring to the educational process. Teachers play a critical role in guiding students, fostering interpersonal relationships, and creating a supportive learning environment.
T6he most effective educational systems will likely be those that leverage the strengths of both AI and human educators. AI will assist by providing real-time, adaptive learning support, while teachers will focus on higher-order tasks such as mentoring, motivating, and inspiring students to think critically and creatively.

\section{Challenges and Future Directions}

AI-based ITS have demonstrated significant advancements in personalized education, yet they face various challenges that must be addressed to realize their full potential. These challenges span areas of scalability, cost, long-term effectiveness, and the ability to handle more complex and creative subjects. Additionally, ethical concerns, integration with traditional classrooms, and accessibility remain key considerations for future developments.

C1. Challenges of Scalability, Cost, and Long-Term Effectiveness
While AI-based ITS offer transformative potential for education, they encounter several practical challenges. Scalability remains a major issue, as deploying these systems across diverse educational contexts, particularly in under-resourced regions, is difficult. Development and deployment costs can be high, limiting widespread adoption. Additionally, there is a growing need to measure the long-term effectiveness of these systems, with rigorous studies required to evaluate not just learning outcomes but also user satisfaction and engagement over time.

C2. Handling Complex, Open-Ended Subjects
Current ITS excel in structured domains like mathematics and grammar but struggle to address complex, open-ended topics that require higher-order thinking. Future ITS should advance their capacity to manage these more nuanced areas to improve the breadth of subjects they can effectively teach.  Most reviewed ITS still struggle to manage open-ended learning tasks that require creativity or ethical reasoning. This limits their applicability in courses where interpretative or divergent thinking is central. Future systems should be co-designed with educators from various disciplines to ensure broader pedagogical alignment.

C3. Enhancing NLP
While many ITS incorporate basic NLP capabilities, significant improvements are necessary to create natural, conversational interactions between systems and students. More advanced NLP could enable systems to better understand student inputs and provide human-like responses, creating a more engaging and effective learning experience.

C4. Addressing Ethical Concerns
As ITS systems collect and analyze vast amounts of student data, privacy and ethical concerns emerge. Future developments must focus on transparent AI decision-making processes and robust data protection measures to ensure ethical use of student information. This will be critical to maintaining trust in these systems, especially as they scale.

C5. Improving Affective Computing
While AI-based ITS can provide personalized feedback, they are still less proficient than human tutors in interpreting students' emotional states. Developing ITS that can recognize and respond to students' affective cues will lead to more empathetic tutoring experiences and improved learning outcomes.

C6. Integrating with Traditional Education
Despite the promise of ITS, their integration into formal educational settings remains limited. Future systems must focus on hybrid models that combine AI-driven personalization with human oversight. This collaboration between AI and human educators will ensure that ITS complement traditional instruction, rather than attempt to replace human teachers.

C7. Expanding Accessibility
To democratize education, AI-based ITS must become more accessible to a broader range of students, particularly those with disabilities or limited access to technology. This includes developing inclusive design practices and addressing the digital divide to ensure that all students benefit from these systems.

C8. Enhancing Cognitive Modeling
ITS currently use AI to model students' cognitive processes, but there is still room for improvement in this area. Enhancing cognitive modeling techniques will allow for more accurate student models and better-tailored instruction, thereby improving learning outcomes.

C9. Developing More Sophisticated Feedback Mechanisms
Providing immediate feedback is a core strength of ITS, but there is potential to develop more nuanced, context-aware feedback systems. These systems could guide students more effectively through complex problem-solving processes, leading to deeper learning experiences.

C10. Improving Long-Term Learning Outcomes
While ITS have been effective in promoting short-term learning gains, more research is needed to assess their impact on long-term knowledge retention and skill development. Future ITS should aim to support not just immediate understanding but also durable, long-term learning outcomes.

C11. Addressing Teacher and Student Acceptance
For ITS to achieve widespread adoption, they must gain the acceptance of both teachers and students. This includes addressing concerns that AI might replace human teachers and clearly demonstrating the benefits of ITS over traditional teaching methods, such as enhanced personalization and data-driven insights.

C12. Pedagogical Challenges
AI-based ITS must align with educational standards and benchmarks to remain relevant in various learning contexts. Additionally, systems should support teachers in integrating AI technologies into traditional classrooms, ensuring that ITS complement rather than disrupt existing pedagogical methods. Engaging students, providing meaningful assessments, and keeping students motivated remain significant challenges that require attention moving forward.

\section{Conclusion}
This review highlights the significant potential of AI-based ITS to transform education through personalized and adaptive learning experiences. The integration of AI techniques such as ML and NLP enables ITS to provide tailored instruction and real-time feedback, enhancing student engagement and learning outcomes.

However, several challenges need to be addressed to fully leverage the benefits of ITS. These include technical issues related to scalability and data integration, ethical concerns about data privacy and algorithmic fairness, and the need for more rigorous experimental designs to evaluate the effectiveness of ITS.

Future research should focus on developing standardized evaluation methods, conducting comparative and longitudinal studies, and exploring the ethical implications of AI in education. By addressing these challenges, researchers and practitioners can enhance the effectiveness and acceptance of ITS in educational settings.

Finally, AI-based ITS represents a promising advancement in educational technology, with the potential to provide personalized, effective, and engaging learning experiences for students worldwide.


\begin{thebibliography}{8}

\bibitem{Dudyrev2021IntelligentVR}
Dudyrev, F., Maksimenkova, O., Mikhailenko, D.: Intelligent Virtual Reality Tutoring Systems as a New Generation of Simulators: Requirements and Opportunities. In: 2021 IEEE Global Engineering Education Conference (EDUCON), pp. 706--718 (2021).

\bibitem{Ali2023ExplainableAI}
Ali, S., Abuhmed, T., El-Sappagh, S., Muhammad, K., Alonso-Moral, J. M., Confalonieri, R., Guidotti, R., Del Ser, J., Díaz-Rodríguez, N., Herrera, F.: Explainable Artificial Intelligence (XAI): What we know and what is left to attain Trustworthy Artificial Intelligence. In: Information Fusion, vol. 99, 101805 (2023).

\bibitem{Koedinger1997Tutoring}
Koedinger, K., Anderson, J. R., Hadley, W. H., Mark, M. A.: Intelligent Tutoring Goes to School in the Big City. Online (1997). \url{https://api.semanticscholar.org/CorpusID:56309126}

\bibitem{Karpouzis2024ExplainableAI}
Karpouzis, K.: Explainable AI for Intelligent Tutoring Systems. In: Farmanbar, M., Tzamtzi, M., Verma, A. K., Chakravorty, A. (eds.) Frontiers of Artificial Intelligence, Ethics, and Multidisciplinary Applications, pp. 59--70. Springer Nature, Singapore (2024).

\bibitem{Smalenberger2022TowardAI}
Smalenberger, M., Smalenberger, K.: Toward Accessible Intelligent Tutoring Systems: Integrating Cognitive Tutors and Conversational Agents. In: International Conference on Artificial Intelligence in Education (2022).

\bibitem{Ramadhan2023ITSGamification}
Ramadhan, M., Warnars, H. L. H. S., Razak, F. H. A.: Combining Intelligent Tutoring Systems and Gamification: A Systematic Literature Review. In: Education and Information Technologies, vol. 29, pp. 6753--6789 (2023).

\bibitem{GarciaBosque2024ChatPLTAI}
Garcia-Bosque, M., Naya-Forcano, A., Cascarosa, E., Aznar, F., Sánchez-Azqueta, C., Celma, S., Aldea, C.: ChatPLT: An intelligent tutoring system for teaching Physics in Higher Education. In: 10th International Conference on Higher Education Advances (HEAd’24) (2024).

\bibitem{CrdovaEsparza2025AIPoweredEA}
Córdova-Esparza, D.-M.: AI-Powered Educational Agents: Opportunities, Innovations, and Ethical Challenges. In: Information (2025).

\bibitem{Gao2025Agent4EduGL}
Gao, W., Liu, Q., Yue, L., Yao, F., Lv, R., Zhang, Z., Wang, H., Huang, Z.: Agent4Edu: Generating Learner Response Data by Generative Agents for Intelligent Education Systems. In: AAAI Conference on Artificial Intelligence (2025).

\bibitem{Wang2025LLMpoweredMF}
Wang, T., Zhan, Y., Lian, J., Hu, Z., Yuan, N. J., Zhang, Q., Xie, X., Xiong, H.: LLM-powered Multi-agent Framework for Goal-oriented Learning in Intelligent Tutoring System. In: Companion Proceedings of the ACM on Web Conference 2025 (2025).

\bibitem{Chan2021VirtualCL}
Chan, P., Van Gerven, T., Dubois, J.-L., Bernaerts, K.: Virtual chemical laboratories: a systematic literature review of research, technologies and instructional design. In: Computers and Education Open (2021).

\bibitem{andes_tech_user_training}
Andes Technology: User Training Downloads and Support. (2024). Accessed: 2024-10-03.

\bibitem{sql_tutor}
SQL Tutor: SQL Tutor Website. (2024). \url{https://sqltutor.de/} Accessed: 2024-10-03.

\bibitem{carnegie_cognitive_tutor}
Carnegie Learning: The Cognitive Tutor: Applying Cognitive Science to Education. (2024). \url{https://www.carnegielearning.com/pages/whitepaper-report/the-cognitive-tutor-applying-cognitive-science-to-education/} Accessed: 2024-10-03.

\bibitem{sim_trainer}
Sim Trainer: Sim Trainer Website. (2024). \url{https://sim-trainer.com/} Accessed: 2024-10-03.

\bibitem{Graesser2005AutoTutorAI}
Graesser, A. C., Chipman, P., Haynes, B. C., Olney, A. M.: AutoTutor: an intelligent tutoring system with mixed-initiative dialogue. In: IEEE Transactions on Education, vol. 48, pp. 612--618 (2005).

\bibitem{BalasoAIEnhancedEP}
Balaso, P., Pawar, L.: AI-Enhanced Education: Personalized Learning and Educational Technology. (Year n/a).

\bibitem{thesims_academy}
Electronic Arts: The Sims Academy. (2024). \url{https://www.thesims.com/fr_FR/the-academy} Accessed: 2024-10-03.

\bibitem{Achieng2022AfricasEP}
Achieng’, R., Wakoli, E., Rodrot, M.: Africa’s Ed-Tech Platforms: Protecting Children’s Right to Privacy. In: Journal of Intellectual Property and Information Technology Law (JIPIT) (2022).

\bibitem{Eduaide2024}
Eduaide: AI-driven Workspace for Dynamic Lessons. (2024). \url{https://www.eduaide.ai/} Accessed: 25 Sept. 2024.

\bibitem{Clancey1987KnowledgebasedTT}
Clancey, W. J.: Knowledge-based tutoring: the GUIDON program. (1987). \url{https://api.semanticscholar.org/CorpusID:61010762}

\bibitem{RoboTutor2024}
Carnegie Mellon University: RoboTutor. (2024). \url{https://www.cmu.edu/scs/robotutor/index.html} Accessed: 24 Sept. 202 ITS with K
\bibitem{PharmaLearn2024}
PharmaLearn: PharmaLearn: Learn from the Experts. (2024). \url{https://learn.pharmalex.com/} Accessed: 24 Sept. 2024.

\bibitem{Viaprolearn2024}
VIAPROLEARN: Road Safety and Driving Education Training. (2024). \url{https://www.viaprolearn.com/} Accessed: 24 Sept. 2024.

\bibitem{Prodiags2024}
Prodiags: Automotive Online Training. (2024). \url{https://prodiags.com/} Accessed: 24 Sept. 2024.

\bibitem{CourseHero2024}
Course Hero: Become an Online Tutor and Earn Money. (2024). \url{https://www.coursehero.com/become-a-tutor/} Accessed: 24 Sept. 2024.

\bibitem{IVIR2024}
Institute for Information Law (IVIR): Summer Course on European Platform Regulation. (2024). \url{https://www.ivir.nl/courses/epr/} Accessed: 24 Sept. 2024.

\bibitem{AutoReturn2024}
AutoReturn: AutoReturn Portal. (2024). \url{https://autoreturn.atlassian.net/} Accessed: 24 Sept. 2024.

\bibitem{Evens1997CIRCSIMTutorAI}
Evens, M. W., Chang, R.-C., Lee, Y. H., Shim, L., Woo, C.-W., Zhang, Y.: CIRCSIM-Tutor: An Intelligent Tutoring System Using Natural Language Dialogue. In: Applied Natural Language Processing Conference (1997).

\bibitem{ ITS with K}
Epic: Epic: With the Patient at the Heart. (2024). \url{https://www.epic.com/} Accessed: 24 Sept. 2024.

\bibitem{MarinerSkills2024}
Mariner Skills: Online Maritime Training. (2024). \url{https://marinerskills.com/} Accessed: 24 Sept. 2024.

\bibitem{FoodSafe2024}
FoodSafe: Food Safety Training and Certification. (2024). \url{https://www.foodsafe.com/} Accessed: 24 Sept. 2024.

\bibitem{IncSys2024}
IncSys: PowerSimulator: Power Grid Simulation Software. (2024). \url{https://www.incsys.com/powersimulator/} Accessed: 24 Sept. 2024.

\bibitem{Gemini2024}
Gemini: Gemini: Cryptocurrency Exchange. (2024). \url{https://www.gemini.com/} Accessed: 24 Sept. 2024.

\bibitem{Simpro2024}
Simpro International: Nitrogen Generators and Power Systems. (2024). \url{https://www.simpro.com/} Accessed: 24 Sept. 2024.

\bibitem{Dart2024}
Dart Communications. (2024). \url{https://www.dart.com/} Accessed: 24 Sept. 2024.

\bibitem{Maestro2024}
Maestro Technologies. (2024). \url{https://maestro.com/} Accessed: 24 Sept. 2024.

\bibitem{Meta2024}
Meta Technologies. (2024). \url{https://www.meta.com/} Accessed: 24 Sept. 2024.

\bibitem{Aerolearn2024}
Aerolearn: Repair Station Training. (2024). \url{https://aerolearn.com/} Accessed: 24 Sept. 2024.

\bibitem{Ersozlu2021MixedRealityLE}
Ersozlu, Z., Ledger, S., Ersozlu, A., Mayne, F., Wildy, H.: Mixed-Reality Learning Environments in Teacher Education: An Analysis of TeachLivE\texttrademark{} Research. In: SAGE Open, vol. 11 (2021).

\bibitem{King2022ChemistryTutor}
King, E. C., Benson, M., Raysor, S., Holme, T. A., Sewall, J., Koedinger, K., Aleven, V., Yaron, D. J.: The Open-Response Chemistry Cognitive Assistance Tutor System: Development and Implementation. In: Journal of Chemical Education, vol. 99, no. 2, pp. 546--552 (2022).

\bibitem{Lee2021ASO}
Lee, B., Park, S. H.: A study on the NCS based curriculum for educating Technical Director for VFX industry with Artificial Intelligence. In: Cartoon and Animation Studies (2021).

\bibitem{HorvthnHadobs2024SOLVEMP}
Horváthné Hadobás, O. E., Stoffova, V.: SOLVE MATH PROBLEMS STEP BY STEP WITH AI SUPPORT. In: INTED Proceedings (2024).

\bibitem{Zerkouk2024AML}
Zerkouk, M., Mihoubi, M., Chikhaoui, B., Wang, S.: A machine learning based model for student’s dropout prediction in online training. In: Education and Information Technologies (2024).

\bibitem{Gao2024LLMenhancedRI}
Gao, J., Chen, B., Zhao, X., Liu, W., Li, X., Wang, Y., Zhang, Z., Wang, W., Ye, Y., Lin, S., Guo, H., Tang, R.: LLM-enhanced Reranking in Recommender Systems. ArXiv, abs/2406.12433 (2024).

\bibitem{AlFaraby2023ReviewON}
Al Faraby, S., Adiwijaya, A., Romadhony, A.: Review on Neural Question Generation for Education Purposes. In: International Journal of Artificial Intelligence in Education, pp. 1--38 (2023).

\bibitem{Santi2022TrainingTF}
Santi, M., Manacero, A., Peronaglio, F. F., Lobato, R. S., Spolon, R., Cavenaghi, M. A.: Training Transformers for Question Generation Task in Intelligent Tutoring Systems. In: 2022 17th Iberian Conference on Information Systems and Technologies (CISTI), pp. 1--6 (2022).

\bibitem{Karpouzis2024BookChapter}
Karpouzis, K.: Explainable AI for Intelligent Tutoring Systems. In: Farmanbar, M. et al. (eds.) Frontiers of Artificial Intelligence, Ethics, and Multidisciplinary Applications, pp. 59--70. Springer Nature Singapore (2024).

\bibitem{Conati2019TowardPX}
Conati, C., Barral, O., Putnam, V., Rieger, L.: Toward personalized XAI: A case study in intelligent tutoring systems. In: Artificial Intelligence, vol. 298, 103503 (2019).

\bibitem{Cutumisu2017AssessingWS}
Cutumisu, M., Blair, K. P., Chin, D. B., Schwartz, D. L.: Assessing Whether Students Seek Constructive Criticism: The Design of an Automated Feedback System for a Graphic Design Task. In: International Journal of Artificial Intelligence in Education, vol. 27, pp. 419--447 (2017).

\bibitem{Jiang2024BeyondAL}
Jiang, Z., Jiang, M.: Beyond Answers: Large Language Model-Powered Tutoring System in Physics Education for Deep Learning and Precise Understanding. ArXiv, abs/2406.10934 (2024).

\bibitem{Crockett2017OnPL}
Crockett, K. A., Latham, A., Whitton, N.: On predicting learning styles in conversational intelligent tutoring systems using fuzzy decision trees. In: International Journal of Human-Computer Studies, vol. 97, pp. 98--115 (2017).

\bibitem{Santhi2013ReviewOI}
Santhi, R., Priya, B., Nandhini, J. M.: Review of intelligent tutoring systems using bayesian approach. ArXiv, abs/1302.7081 (2013).

\bibitem{Ingkavara2023TrendsOA}
Ingkavara, T., Wongkia, W., Panjaburee, P.: Trends of Adaptive/Personalized Learning and Intelligent Tutoring Systems in Mathematics: A Review of Academic Publications from 2010 to 2022. In: 2023 IEEE 5th Eurasia Conference on Biomedical Engineering, Healthcare and Sustainability (2023).

\bibitem{Alam2023HarnessingAI}
Alam, A.: Harnessing the Power of AI to Create Intelligent Tutoring Systems for Enhanced Classroom Experience and Improved Learning Outcomes. In: Intelligent Communication Technologies and Virtual Mobile Networks, pp. 571--591. Springer Nature Singapore (2023).

\bibitem{Aleven2009ANP}
Aleven, V., McLaren, B. M., Sewall, J., Koedinger, K.: A New Paradigm for Intelligent Tutoring Systems: Example-Tracing Tutors. In: International Journal of Artificial Intelligence in Education, vol. 19, pp. 105--154 (2009).

\bibitem{Page2020TheP2}
Page, M. J., McKenzie, J. E., Bossuyt, P. M. M., et al.: The PRISMA 2020 statement: an updated guideline for reporting systematic reviews. In: Systematic Reviews, vol. 10 (2020).

\bibitem{Abbas2023RoleOA}
Abbas, N., Ali, I., Manzoor, R., Hussain, T., Hussain, M. H. A.: Role of Artificial Intelligence Tools in Enhancing Students' Educational Performance at Higher Levels. (2023).

\bibitem{Budgen2006PerformingSL}
Budgen, D., Brereton, P.: Performing systematic literature reviews in software engineering. In: Proceedings of the 28th International Conference on Software Engineering (2006).

\bibitem{FernndezHerrero2024EvaluatingRA}
Fernández-Herrero, J.: Evaluating Recent Advances in Affective Intelligent Tutoring Systems: A Scoping Review of Educational Impacts and Future Prospects. In: Education Sciences (2024).

\bibitem{Weaver2024ThePO}
Weaver, J. C., Matangula, T. C., Matney, G. T.: The power of feedback in teacher education. In: International Journal for Lesson \& Learning Studies (2024).

\bibitem{Qwaider2018ExcelIT}
Qwaider, S. R., Abu-Naser, S. S.: Excel Intelligent Tutoring System. In: Proceedings (2018).

\bibitem{Huang2020TheIO}
Huang, R.-T., Ritzhaupt, A. D., Sommer, M., Zhu, J., Stephen, A., Valle, N., Hampton, J., Li, J.: The impact of gamification in educational settings on student learning outcomes: a meta-analysis. In: Educational Technology Research and Development, vol. 68, pp. 1875--1901 (2020).

\bibitem{Andrade2016TheBA}
Andrade, F. R. H., Mizoguchi, R., Isotani, S.: The Bright and Dark Sides of Gamification. In: International Conference on Intelligent Tutoring Systems (2016).

\bibitem{Alsawaier2018TheEO}
Alsawaier, R. S.: The effect of gamification on motivation and engagement. In: Proceedings (2018).

\bibitem{FerreiradaRocha2023GamificationAO}
Ferreira da Rocha, F. D., Lemos, B., de Brito, P. H., Santos, R., Rodrigues, L., Isotani, S., Dermeval, D.: Gamification and open learner model: An experimental study on the effects on self-regulatory learning characteristics. In: Education and Information Technologies, vol. 29, pp. 3525--3546 (2023).

\bibitem{Shih2023MathematicsIT}
Shih, S.-C., Chang, C.-C., Kuo, B.-C., Huang, Y.-H.: Mathematics intelligent tutoring system for learning multiplication and division of fractions based on diagnostic teaching. In: Education and Information Technologies, pp. 1--22 (2023).

\bibitem{UriartePortillo2023IntelligentAR}
Uriarte-Portillo, A., Zataráin-Cabada, R., Barrón Estrada, M. L., Ibáñez, M. B., González-Barrón, L.-M.: Intelligent Augmented Reality for Learning Geometry. In: Information, vol. 14, 245 (2023).

\bibitem{Graesser2018ElectronixTutorAI}
Graesser, A. C., Hu, X., Nye, B. D., VanLehn, K., Kumar, R., Heffernan, C., Heffernan, N. T., Woolf, B. P., Olney, A. M., Rus, V., et al.: ElectronixTutor: an intelligent tutoring system with multiple learning resources for electronics. In: International Journal of STEM Education, vol. 5 (2018).

\bibitem{Kurni2023}
Kurni, M., Mohammed, M. S., Srinivasa, K. G.: Intelligent Tutoring Systems. In: A Beginner's Guide to Introduce Artificial Intelligence in Teaching and Learning, pp. 29--44. Springer, Cham (2023).

\bibitem{Tang2023CombiningGA}
Tang, Y., Hare, R.: Combining Gamification and Intelligent Tutoring Systems in a Serious Game for Engineering Education. ArXiv, abs/2305.16568 (2023).

\bibitem{Makkubhai2023EnhancingEP}
Makkubhai, I. M., Afreen, M., Makkubhai, J. I., B. A. Final, Year Student: Enhancing Educational Pedagogy through Intelligent Systems: Harnessing Artificial Intelligence for Advancing Progressive Teaching Practices. In: Proceedings (2023).

\bibitem{VanLehn2010TheAP}
VanLehn, K., van de Sande, B., Shelby, R., Gershman, S.: The Andes Physics Tutoring System: An Experiment in Freedom. In: Advances in Intelligent Tutoring Systems (2010).

\bibitem{Wang1997SQLTA}
Wang, H.: SQL Tutor+: A co-operative ITS with repository support. In: Information and Software Technology, vol. 39, pp. 343--350 (1997).

\bibitem{Anderson1997ACTRAT}
Anderson, J. R., Matessa, M., Lebiere, C.: ACT-R: A Theory of Higher Level Cognition and Its Relation to Visual Attention. In: Human-Computer Interaction, vol. 12, pp. 439--462 (1997).

\bibitem{Bradc2022DesignOA}
Bradač, V., Smolka, P., Kotyrba, M., Průdek, T.: Design of an Intelligent Tutoring System to Create a Personalized Study Plan Using Expert Systems. In: Applied Sciences (2022).

\bibitem{Frasson1992FromES}
Frasson, C.: From Expert Systems to Intelligent Tutoring Systems. In: Proceedings (1992).

\bibitem{Gonzlez2013DesigningIT}
González, C., Burguillo, J. C., Llamas, M., Laza, R.: Designing Intelligent Tutoring Systems: A Personalization Strategy using Case-Based Reasoning and Multi-Agent Systems. In: Proceedings (2013).

\bibitem{Yan2024ARO}
Yan, A., Cheng, Z.: A Review of the Development and Future Challenges of Case-Based Reasoning. In: Applied Sciences (2024).

\bibitem{Perner2019CaseBasedR}
Perner, P.: Case-Based Reasoning - Methods, Techniques, and Applications. In: Iberoamerican Congress on Pattern Recognition (2019).

\bibitem{Reyes2001AgentRA}
Reyes, R. L., Sison, R. C.: Agent Representation and Communication in CBR-Tutor. In: International Conference on Health Information Science (2001).

\bibitem{Piro2024MyLearningTalkAL}
Piro, L., Bianchi, T., Alessandrelli, L., Chizzola, A., Casiraghi, D., Sancassani, S., Gatti, N.: MyLearningTalk: An LLM-Based Intelligent Tutoring System. In: International Conference on Web Engineering (2024).

\bibitem{Algherairy2023ARO}
Algherairy, A., Ahmed, M.: A review of dialogue systems: current trends and future directions. In: Neural Computing and Applications (2023).

\bibitem{Feng2024CourseAssistPA}
Feng, T., Liu, S., Ghosal, D.: CourseAssist: Pedagogically Appropriate AI Tutor for Computer Science Education. ArXiv, abs/2407.10246 (2024).

\bibitem{Chughtai2015}
Chughtai, R., Zhang, S., Craig, S. D.: Usability Evaluation of Intelligent Tutoring System: ITS from a Usability Perspective. In: Proceedings of the Human Factors and Ergonomics Society Annual Meeting, vol. 59, no. 1, pp. 367--371 (2015).

\bibitem{Wang2021UsabilityOA}
Wang, T.-H., Lin, H.-C. K., Chen, H.-R., Huang, Y.-M., Yeh, W.-T., Li, C.-T.: Usability of an Affective Emotional Learning Tutoring System for Mobile Devices. In: Sustainability (2021).

\bibitem{Lin2022EyeMA}
Lin, H.-C. K., Liao, Y.-C., Wang, H.-T.: Eye Movement Analysis and Usability Assessment on Affective Computing Combined with Intelligent Tutoring System. In: Sustainability (2022).

\bibitem{Graesser2006AutoTutorAC}
Graesser, A. C., Olney, A. M., Haynes, B. C., Chipman, P.: AutoTutor: A Cognitive System That Simulates a Tutor Through Mixed-Initiative Dialogue. In: Proceedings (2006).

\bibitem{ZataranCabada2014EmotionRI}
Zataráin-Cabada, R., Barrón Estrada, M. L., Alor-Hernández, G., Reyes García, C. A.: Emotion Recognition in Intelligent Tutoring Systems for Android-Based Mobile Devices. In: Mexican International Conference on Artificial Intelligence (2014).

\bibitem{Xu2018LearningEE}
Xu, T., Zhou, Y., Wang, Z., Peng, Y.: Learning Emotions EEG-based Recognition and Brain Activity: A Survey Study on BCI for Intelligent Tutoring System. In: ANT/SEIT (2018).

\bibitem{Chen2019ResearchOI}
Chen, Y., Zhang, Y.: Research on Intelligent Tutoring System Based on Data-Mining Algorithms. In: 2019 International Conference on Intelligent Transportation, Big Data \& Smart City (ICITBS), pp. 443--446 (2019).

\bibitem{Lin2023ArtificialII}
Lin, C.-C., Huang, A. Y. Q., Lu, O. H. T.: Artificial intelligence in intelligent tutoring systems toward sustainable education: a systematic review. In: Smart Learning Environments, vol. 10, pp. 1--22 (2023).

\bibitem{Thomas2023ImprovingSL}
Thomas, D. R., Lin, J., Gatz, E., Gurung, A., Gupta, S., Norberg, K., Fancsali, S. E., Aleven, V., Branstetter, L., Brunskill, E., Koedinger, K.: Improving Student Learning with Hybrid Human-AI Tutoring: A Three-Study Quasi-Experimental Investigation. In: Proceedings of the 14th Learning Analytics and Knowledge Conference (2023).

\bibitem{VanLehn2011TheRE}
VanLehn, K.: The Relative Effectiveness of Human Tutoring, Intelligent Tutoring Systems, and Other Tutoring Systems. In: Educational Psychologist, vol. 46, pp. 197--221 (2011).

\bibitem{Chi2011AnEO}
Chi, M., VanLehn, K., Litman, D. J., Jordan, P. W.: An Evaluation of Pedagogical Tutorial Tactics for a Natural Language Tutoring System: A Reinforcement Learning Approach. In: International Journal of Artificial Intelligence in Education, vol. 21, pp. 83--113 (2011).

\bibitem{Shermis2003AutomatedES}
Shermis, M. D., Burstein, J.: Automated Essay Scoring: A Cross-disciplinary Perspective. In: Proceedings (2003).

\bibitem{Feng2009UsingMM}
Feng, M., Heffernan, N. T., Heffernan, C., Mani, M.: Using Mixed-Effects Modeling to Analyze Different Grain-Sized Skill Models in an Intelligent Tutoring System. In: IEEE Transactions on Learning Technologies, vol. 2, pp. 79--92 (2009).

\bibitem{Phobun2010AdaptiveIT}
Phobun, P., Vicheanpanya, J.: Adaptive intelligent tutoring systems for e-learning systems. In: Procedia - Social and Behavioral Sciences, vol. 2, pp. 4064--4069 (2010).

\bibitem{Ifenthaler2024ArtificialII}
Ifenthaler, D., Majumdar, R., Gorissen, P., Judge, M., Mishra, S., Raffaghelli, J., Shimada, A.: Artificial Intelligence in Education: Implications for Policymakers, Researchers, and Practitioners. In: Technology, Knowledge and Learning (2024).

\bibitem{Adel2024TheCO}
Adel, A.: The Convergence of Intelligent Tutoring, Robotics, and IoT in Smart Education for the Transition from Industry 4.0 to 5.0. In: Smart Cities (2024).

\bibitem{Laak2024AIAP}
Laak, K.-J., Aru, J.: AI and personalized learning: bridging the gap with modern educational goals. ArXiv, abs/2404.02798 (2024).

\bibitem{Hollan1984STEAMERAI}
Hollan, J., Hutchins, E. L., Weitzman, L.: STEAMER: An Interactive Inspectable Simulation-Based Training System. In: AI Magazine, vol. 5, pp. 15--27 (1984).

\bibitem{Katz2020Sherlock2A}
Katz, S., Lesgold, A. M., Hughes, E. J., Peters, D., Eggan, G., Gordin, M., Greenberg, L.: Sherlock 2: An Intelligent Tutoring System Built on the LRDC Tutor Framework. (2020).

\bibitem{Kamruzzaman2023AIAI}
Kamruzzaman, M. M., Alanazi, S. A., Alruwaili, M., Alshammari, N. O., Elaiwat, S., Abu-Zanona, M., Innab, N., Elzaghmouri, B. M., Alanazi, B. A.: AI- and IoT-Assisted Sustainable Education Systems during Pandemics, such as COVID-19, for Smart Cities. In: Sustainability (2023).

\bibitem{Bilad2023RecentPI}
Bilad, M. R., Yaqin, L. N., Zubaidah, S.: Recent Progress in the Use of Artificial Intelligence Tools in Education. In: Jurnal Penelitian dan Pengkajian Ilmu Pendidikan: e-Saintika, vol. 8, pp. 51--57 (2024).

\bibitem{Isaiah2015RuleBM}
Isaiah, A. I.: Rule Based Multi Agent Student Modeling Intelligent Tutoring System. In: Proceedings (2015).

\bibitem{Aamodt1994CaseBasedRF}
Aamodt, A., Plaza, E.: Case-Based Reasoning: Foundational Issues, Methodological Variations, and System Approaches. In: AI Communications, vol. 7, pp. 39--59 (1994).

\bibitem{Koedinger1997IntelligentTG}
Koedinger, K., Anderson, J. R., Hadley, W. H., Mark, M. A.: Intelligent Tutoring Goes To School in the Big City. In: Proceedings (1997).

\bibitem{Litman2004ITSPOKEAI}
Litman, D. J., Silliman, S.: ITSPOKE: An Intelligent Tutoring Spoken Dialogue System. In: NAACL (2004).

\bibitem{Brusilovsky2007UserMF}
Brusilovsky, P., Millán, E.: User Models for Adaptive Hypermedia and Adaptive Educational Systems. In: The Adaptive Web (2007).

\bibitem{Sousa2022AIAB}
Sousa, M. J., Dal Mas, F., Gonçalves, S. P., Calandra, D. M.: AI and Blockchain as New Triggers in the Education Arena. In: European Journal of Investigation in Health, Psychology and Education, vol. 12, pp. 445--447 (2022). \url{https://api.semanticscholar.org/CorpusID:248060133}

\bibitem{Paredes2004AMA}
Paredes, P., Rodríguez, P.: A Mixed Approach to Modelling Learning Styles in Adaptive Educational Hypermedia. In: Advanced Technology for Learning, vol. 1 (2004).

\bibitem{Manouselis2012RecommenderSF}
Manouselis, N., Drachsler, H., Verbert, K., Duval, E.: Recommender Systems for Learning. In: Springer Briefs in Electrical and Computer Engineering (2012).

\bibitem{Corbett2005KnowledgeTM}
Corbett, A. T., Anderson, J. R.: Knowledge tracing: Modeling the acquisition of procedural knowledge. In: User Modeling and User-Adapted Interaction, vol. 4, pp. 253--278 (2005).

\bibitem{DMello2012AutoTutorAA}
D’Mello, S. K., Graesser, A. C.: AutoTutor and affective autotutor: Learning by talking with cognitively and emotionally intelligent computers that talk back. In: ACM Transactions on Interactive Intelligent Systems, vol. 2, 23:1--23:39 (2012).

\bibitem{Anderson1995CognitiveTL}
Anderson, J. R., Corbett, A. T., Koedinger, K., Pelletier, R.: Cognitive Tutors: Lessons Learned. In: The Journal of the Learning Sciences, vol. 4, pp. 167--207 (1995).

\bibitem{deJong1998ScientificDL}
de Jong, T., van Joolingen, W. R.: Scientific Discovery Learning with Computer Simulations of Conceptual Domains. In: Review of Educational Research, vol. 68, pp. 179--201 (1998).

\bibitem{Laine2022SystematicRO}
Laine, J., Lindqvist, T., Korhonen, T. S., Hakkarainen, K.: Systematic Review of Intelligent Tutoring Systems for Hard Skills Training in Virtual Reality Environments. In: International Journal of Technology in Education and Science (2022).

\bibitem{Nesbit2015WorkIP}
Nesbit, J. C., Liu, L., Liu, Q., Adesope, O. O.: Work in Progress: Intelligent Tutoring Systems in Computer Science and Software Engineering Education. (2015).

\bibitem{Li2021ABC}
Li, Z., Ma, Z.: A blockchain-based credible and secure education experience data management scheme supporting for searchable encryption. In: China Communications, vol. 18, pp. 172--183 (2021).

\bibitem{Lan2024SurveyON}
Lan, Y., Li, X., Du, H., Lu, X., Gao, M., Qian, W., Zhou, A.: Survey of Natural Language Processing for Education: Taxonomy, Systematic Review, and Future Trends. ArXiv, abs/2401.07518 (2024).

\bibitem{Vindigni2023AdaptiveAR}
Vindigni, G.: Adaptive and Re-adaptive Pedagogies in Higher Education: A Comparative, Longitudinal Study of Their Impact on Professional Competence Development across Diverse Curricula. In: European Journal of Theoretical and Applied Sciences (2023).

\bibitem{Muangprathub2020LearningRW}
Muangprathub, J., Boonjing, V., Chamnongthai, K.: Learning recommendation with formal concept analysis for intelligent tutoring system. In: Heliyon, vol. 6 (2020).

\bibitem{AlNakhal2017AdaptiveIT}
Al-Nakhal, M. A., Abu Naser, S. S.: Adaptive Intelligent Tutoring System for learning Computer Theory. (2017).

\bibitem{Hooshyar2016ApplyingAO}
Hooshyar, D., Ahmad, R. B., Yousefi, M., Fathi, M., Horng, S.-J., Lim, H.: Applying an online game-based formative assessment in a flowchart-based intelligent tutoring system for improving problem-solving skills. In: Computers \& Education, vol. 94, pp. 18--36 (2016).

\bibitem{SteenbergenHu2014AMO}
Steenbergen-Hu, S., Cooper, H.: A meta-analysis of the effectiveness of intelligent tutoring systems on college students' academic learning. In: Journal of Educational Psychology, vol. 106, pp. 331--347 (2014).

\bibitem{SteenbergenHu2013AMO}
Steenbergen-Hu, S., Cooper, H.: A meta-analysis of the effectiveness of intelligent tutoring systems on K–12 students’ mathematical learning. In: Journal of Educational Psychology, vol. 105, pp. 970--987 (2013).

\bibitem{Nkambou2023LearningLR}
Nkambou, R., Brisson, J., Nyamen Tato, A. A., Robert, S.: Learning Logical Reasoning Using an Intelligent Tutoring System: A Hybrid Approach to Student Modeling. In: AAAI Conference on Artificial Intelligence (2023).

\bibitem{Binh2021ResponsiveSM}
Binh, H. T., Trung, N. Q., Duy, B. T.: Responsive student model in an intelligent tutoring system and its evaluation. In: Education and Information Technologies, vol. 26, pp. 4969--4991 (2021).

\bibitem{Wang2023ExaminingTA}
Wang, H., Tlili, A., Huang, R., Cai, Z., Li, M., Cheng, Z., Yang, D., Li, M., Zhu, X., Fei, C.: Examining the applications of intelligent tutoring systems in real educational contexts: A systematic literature review from the social experiment perspective. In: Education and Information Technologies, pp. 1--36 (2023).

\bibitem{Chughtai2015UsabilityEO}
Chughtai, R., Zhang, S., Craig, S. D.: Usability evaluation of intelligent tutoring system. In: Proceedings of the Human Factors and Ergonomics Society Annual Meeting, vol. 59, pp. 367--371 (2015).

\bibitem{Swiecki2022AssessmentIT}
Swiecki, Z., Khosravi, H., Chen, G., Martínez Maldonado, R., Lodge, J. M., Milligan, S., Selwyn, N., Gašević, D.: Assessment in the age of artificial intelligence. In: Computers and Education: Artificial Intelligence, vol. 3, 100075 (2022).

\bibitem{Heffernan2000IntelligentTS}
Heffernan, N. T., Koedinger, K.: Intelligent Tutoring Systems are Missing the Tutor: Building a More Strategic Dialog-Based Tutor. Online (2000). \url{https://api.semanticscholar.org/CorpusID:62663223}

\bibitem{Gonzlez2006ACA}
González, C., Burguillo, J. C., Llamas, M.: A Case-Based Approach for Building Intelligent Tutoring Systems. In: 2006 7th International Conference on Information Technology Based Higher Education and Training, pp. 442--446 (2006).

\bibitem{Walkington2013UsingAL}
Walkington, C. A.: Using adaptive learning technologies to personalize instruction to student interests: The impact of relevant contexts on performance and learning outcomes. In: Journal of Educational Psychology, vol. 105, pp. 932--945 (2013).

\bibitem{Bellarhmouch2022APA}
Bellarhmouch, Y., Jeghal, A., Tairi, H., Benjelloun, N.: A proposed architectural learner model for a personalized learning environment. In: Education and Information Technologies, vol. 28, pp. 4243--4263 (2022).

\bibitem{Abdi2023StudentsFA}
Abdi, A., Sedrakyan, G., Veldkamp, B. P., van Hillegersberg, J., van den Berg, S. M.: Students feedback analysis model using deep learning-based method and linguistic knowledge for intelligent educational systems. In: Soft Computing, vol. 27, pp. 14073--14094 (2023).

\bibitem{Hasan2020TheTF}
Hasan, M. A., Noor, N. F. M., Rahman, S. S. A., Rahman, M. M.: The Transition From Intelligent to Affective Tutoring System: A Review and Open Issues. In: IEEE Access, vol. 8, pp. 204612--204638 (2020).

\bibitem{Mousavinasab2018IntelligentTS}
Mousavinasab, E., Zarifsanaiey, N., Niakan Kalhori, S. R., Rakhshan, M., Keikha, L., Ghazi Saeedi, M.: Intelligent tutoring systems: a systematic review of characteristics, applications, and evaluation methods. In: Interactive Learning Environments, vol. 29, pp. 142--163 (2018).

\bibitem{Djunaidi2023CharacteristicsAA}
Djunaidi, K., Siswipraptini, P. C., Sikumbang, H.: Characteristics and Application of Intelligent Tutoring Systems: A Review. In: 2023 10th International Conference on ICT for Smart Society (ICISS), pp. 1--7 (2023).

\bibitem{CastroSchez2020AnIT}
Castro-Schez, J. J., González-Morcillo, C., Albusac, J., Vallejo-Fernandez, D.: An intelligent tutoring system for supporting active learning: A case study on predictive parsing learning. In: Information Sciences, vol. 544, pp. 446--468 (2020).

\bibitem{Ou2024TransformingET}
Ou, S.: Transforming Education: The Evolving Role of Artificial Intelligence in The Students Academic Performance. In: International Journal of Education and Humanities (2024).

\bibitem{Mainetti2022DigitalBE}
Mainetti, L., Paiano, R., Pedone, M., Quarta, M., Dervishi, E.: Digital Brick: Enhancing the Student Experience Using Blockchain, Open Badges and Recommendations. In: Education Sciences (2022).

\bibitem{Terzi2021ALL}
Terzi, S., Stamelos, I., Votis, K., Tsiatsos, T.: A Life-Long Learning Education Passport Powered by Blockchain Technology and Verifiable Digital Credentials: The BlockAdemiC Project. In: SEFM Workshops (2021). \url{https://api.semanticscholar.org/CorpusID:252546481}

\bibitem{Guo2021EvolutionAT}
Guo, L., Wang, D., Gu, F., Li, Y., Wang, Y., Zhou, R.: Evolution and trends in intelligent tutoring systems research: a multidisciplinary and scientometric view. In: Asia Pacific Education Review, vol. 22, pp. 441--461 (2021).

\bibitem{Alsobhi2023BlockchainbasedMS}
Alsobhi, H., Alakhtar, R. A., Ubaid, A., Hussain, O. K., Hussain, F. K.: Blockchain-based micro-credentialing system in higher education institutions: Systematic literature review. In: Knowledge-Based Systems, vol. 265, 110238 (2023).

\bibitem{Kumar2019AnAF}
Kumar, A., Ahuja, N. J.: An Adaptive Framework of Learner Model Using Learner Characteristics for Intelligent Tutoring Systems. In: Advances in Intelligent Systems and Computing (2019).

\bibitem{noodlefactory}
Noodle Factory. (2024). \url{https://www.noodlefactory.ai/} Accessed: 2024-07-25.

\bibitem{tutello}
Tutello. (2024). \url{https://www.tutello.com/} Accessed: 2024-07-25.

\bibitem{Ermit2020DesignFO}
Erümit, A. K., Çetin, İ.: Design framework of adaptive intelligent tutoring systems. In: Education and Information Technologies, vol. 25, pp. 4477--4500 (2020).

\bibitem{Fang2018AMO}
Fang, Y., Ren, Z., Hu, X., Graesser, A. C.: A meta-analysis of the effectiveness of ALEKS on learning. In: Educational Psychology, vol. 39, pp. 1278--1292 (2018).

\bibitem{Feng2021ASR}
Feng, S., Magana, A. J., Kao, D.: A Systematic Review of Literature on the Effectiveness of Intelligent Tutoring Systems in STEM. In: 2021 IEEE Frontiers in Education Conference (FIE), pp. 1--9 (2021).

\bibitem{Tricco2016ASR}
Tricco, A. C., Lillie, E., Zarin, W., et al.: A scoping review on the conduct and reporting of scoping reviews. In: BMC Medical Research Methodology, vol. 16 (2016).

\bibitem{Nkambou2010AdvancesII}
Nkambou, R., Mizoguchi, R., Bourdeau, J.: Advances in Intelligent Tutoring Systems. (2010).

\bibitem{VanLehn2006TheBO}
VanLehn, K.: The Behavior of Tutoring Systems. In: International Journal of Artificial Intelligence in Education, vol. 16, pp. 227--265 (2006).

\bibitem{Shute2007FocusOF}
Shute, V. J.: Focus on Formative Feedback. In: Review of Educational Research, vol. 78, pp. 153--189 (2007).

\bibitem{Dzikovska2014BEETLEID}
Dzikovska, M. O., Steinhauser, N. B., Farrow, E., Moore, J. D., Campbell, G. E.: BEETLE II: Deep Natural Language Understanding and Automatic Feedback Generation for Intelligent Tutoring in Basic Electricity and Electronics. In: International Journal of Artificial Intelligence in Education, vol. 24, pp. 284--332 (2014).

\bibitem{Paladines2020ASL}
Paladines, J., Ramírez, J.: A Systematic Literature Review of Intelligent Tutoring Systems With Dialogue in Natural Language. In: IEEE Access, vol. 8, pp. 164246--164267 (2020).

\bibitem{Modi2022IncorporatingFT}
Modi, B. A.: Incorporating Focus to Enhance Staff-student Interactions in Formative Feedback. In: Proceedings of the 27th ACM Conference on Innovation and Technology in Computer Science Education Vol. 2 (2022).

\bibitem{Sindi2005AML}
Sindi, H. F.: A machine learning approach for intelligent tutoring systems. (2005).

\bibitem{Jeremic2012StudentMA}
Jeremić, Z., Jovanović, J., Gašević, D.: Student modeling and assessment in intelligent tutoring of software patterns. In: Expert Systems with Applications, vol. 39, pp. 210--222 (2012).

\bibitem{Gong2014StudentMI}
Gong, Y.-T.: Student Modeling in Intelligent Tutoring Systems. (2014). \url{https://api.semanticscholar.org/CorpusID:108388149}

\bibitem{Kass1989StudentMI}
Kass, R.: Student Modeling in Intelligent Tutoring Systems — Implications for User Modeling. (1989).

\bibitem{Schmucker2022TransferableSP}
Schmucker, R., Mitchell, T. M.: Transferable Student Performance Modeling for Intelligent Tutoring Systems. ArXiv, abs/2202.03980 (2022).

\bibitem{Alday2018BayesianNI}
Alday, R. B.: Bayesian Networks in Intelligent Tutoring Systems as an Assessment of Student Performance using Student Modeling. In: Proceedings of the 2nd International Conference on Algorithms, Computing and Systems (2018).

\bibitem{Yang2021MachineLS}
Yang, C., Chiang, F.-K., Cheng, Q., Ji, J.: Machine Learning-Based Student Modeling Methodology for Intelligent Tutoring Systems. In: Journal of Educational Computing Research, vol. 59, pp. 1015--1035 (2021).

\bibitem{GonzlezCalatayud2021ArtificialIF}
González-Calatayud, V., Prendes-Espinosa, P., Roig-Vila, R.: Artificial Intelligence for Student Assessment: A Systematic Review. In: Applied Sciences (2021).

\bibitem{Santos2024OpportunitiesAC}
dos Santos, S. C., Junior, G. A. S.: Opportunities and Challenges of AI to Support Student Assessment in Computing Education: A Systematic Literature Review. In: International Conference on Computer Supported Education (2024).

\bibitem{Kulshreshtha2022FewshotQG}
Kulshreshtha, D., Shayan, M., Belfer, R., Reddy, S., Serban, I., Kochmar, E.: Few-shot Question Generation for Personalized Feedback in Intelligent Tutoring Systems. In: PAIS@ECAI (2022).

\bibitem{Bai2022AutomatedES}
Bai, J. Y. H., Zawacki-Richter, O., Bozkurt, A., Lee, K., Fanguy, M., Sari, B. C., Marín, V. I.: Automated Essay Scoring (AES) Systems: Opportunities and Challenges for Open and Distance Education. In: Tenth Pan-Commonwealth Forum on Open Learning (2022).

\bibitem{Badshah2023TowardsSE}
Badshah, A., Ghani, A., Daud, A., Jalal, A., Bilal, M., Crowcroft, J. A.: Towards Smart Education through Internet of Things: A Survey. In: ACM Computing Surveys, vol. 56, pp. 1--33 (2023).

\bibitem{Mokmin2017TheDA}
Mokmin, N. A. M., Masood, M.: The design and development of an intelligent tutoring system as a part of the architecture of internet of things (IoT). In: International Conference on Telecommunications and Communication Engineering (2017).

\bibitem{Frank2024LeveragingGF}
Frank, L., Herth, F., Stuwe, P., Klaiber, M., Gerschner, F., Theissler, A.: Leveraging GenAI for an Intelligent Tutoring System for R: A Quantitative Evaluation of Large Language Models. In: 2024 IEEE Global Engineering Education Conference (EDUCON), pp. 1--9 (2024).

\bibitem{Yu2023GenerativeAI}
Yu, H., Guo, Y.: Generative artificial intelligence empowers educational reform: current status, issues, and prospects. In: Frontiers in Education (2023).

\bibitem{Cal2024TowardsET}
Calò, T., MacLellan, C. J.: Towards Educator-Driven Tutor Authoring: Generative AI Approaches for Creating Intelligent Tutor Interfaces. In: ACM Conference on Learning @ Scale (2024).

\bibitem{Akgun2021ArtificialII}
Akgun, S., Greenhow, C.: Artificial intelligence in education: Addressing ethical challenges in K-12 settings. In: AI and Ethics, vol. 2, pp. 431--440 (2021).

\bibitem{Airaj2024EthicalAI}
Airaj, M.: Ethical artificial intelligence for teaching-learning in higher education. In: Education and Information Technologies (2024).

\bibitem{Khatri2023ArtificialI}
Khatri, B. B., Karki, P. D.: Artificial Intelligence (AI) in Higher Education: Growing Academic Integrity and Ethical Concerns. In: Nepalese Journal of Development and Rural Studies (2023).




\bibitem{article1}
Govea, J., Ocampo, E., Revelo-Tapia, S., Villegas, W.: Optimization and Scalability of Educational Platforms: Integration of Artificial Intelligence and Cloud Computing. In: Computers, vol. 12, p. 223 (2023). doi:10.3390/computers12110223











\bibitem{Abazi2024ProsCons}
Abazi Chaushi, B., Ismaili, F., Chaushi, A.: Pros and Cons of Artificial Intelligence in Education. In: International Journal of Advanced Natural Sciences and Engineering Researches, vol. 8, pp. 51--57 (2024).

\bibitem{Govea2023Optimization}
Govea, J., Ocampo, E., Revelo-Tapia, S., Villegas, W.: Optimization and Scalability of Educational Platforms: Integration of Artificial Intelligence and Cloud Computing. In: Computers, vol. 12, 223 (2023). doi:10.3390/computers12110223

\bibitem{Peres2020IndustrialAI}
Peres, R. S., Jia, X., Lee, J., Sun, K., Colombo, A. W., Barata, J.: Industrial Artificial Intelligence in Industry 4.0 - Systematic Review, Challenges and Outlook. In: IEEE Access, vol. 8, pp. 220121--220139 (2020).

\bibitem{Chaudhry2021ArtificialII}
Chaudhry, M. A., Kazim, E.: Artificial Intelligence in Education (AIEd): a high-level academic and industry note 2021. In: AI and Ethics, vol. 2, pp. 157--165 (2021).

\bibitem{Koedinger2013NewPF}
Koedinger, K., Brunskill, E., Baker, R., McLaughlin, E., Stamper, J. C.: New Potentials for Data-Driven Intelligent Tutoring System Development and Optimization. In: AI Magazine, vol. 34, pp. 27--41 (2013).

\bibitem{Stamper2024EnhancingLF}
Stamper, J., Xiao, R., Hou, X.: Enhancing LLM-Based Feedback: Insights from Intelligent Tutoring Systems and the Learning Sciences. ArXiv, abs/2405.04645 (2024).

\bibitem{Ramadhan2023CombiningIT}
Ramadhan, A., Warnars, H. L. H. S., Razak, F. H. A.: Combining intelligent tutoring systems and gamification: a systematic literature review. In: Education and Information Technologies, vol. 29, pp. 6753--6789 (2023).

\bibitem{Liu2024TheDO}
Liu, H., Zhang, Y., Jia, J.: The Design of Guiding and Adaptive Prompts for Intelligent Tutoring Systems and Its Effect on Students’ Mathematics Learning. In: IEEE Transactions on Learning Technologies, vol. 17, pp. 1379--1389 (2024).

\bibitem{Heilman2006LanguageLC}
Heilman, M.: Language Learning: Challenges for Intelligent Tutoring Systems. (2006).

\bibitem{Bernacki2021ASR}
Bernacki, M. L., Greene, M. J., Lobczowski, N. G.: A Systematic Review of Research on Personalized Learning: Personalized by Whom, to What, How, and for What Purpose(s)? In: Educational Psychology Review, vol. 33, pp. 1675--1715 (2021).

\bibitem{Mislevy2020AutomatedSI}
Mislevy, R. J., Yan, D., Gobert, J. D., Sao Pedro, M. A.: Automated Scoring in Intelligent Tutoring Systems. (2020).

\bibitem{Hussein2019AutomatedLE}
Hussein, M. A., Hassan, H. A., Nassef, M.: Automated language essay scoring systems: a literature review. In: PeerJ Computer Science, vol. 5 (2019).

\bibitem{Graesser2001IntelligentTS}
Graesser, A. C., VanLehn, K., Rosé, C. P., Jordan, P. W., Harter, D.: Intelligent Tutoring Systems with Conversational Dialogue. In: AI Magazine, vol. 22, pp. 39--52 (2001).

\bibitem{Bb2022UsingSA}
Bóbo, M. L. D. R., Campos, F., Stroele, V., David, J. M. N., Braga, R. I., Torrent, T. T.: Using Sentiment Analysis to Identify Student Emotional State to Avoid Dropout in E-Learning. In: International Journal of Distance Education Technologies, vol. 20, pp. 1--24 (2022).

\bibitem{Sychev2024EducationalMF}
Sychev, O.: Educational models for cognition: Methodology of modeling intellectual skills for intelligent tutoring systems. In: Cognitive Systems Research, vol. 87 (2024).

\bibitem{SotoForero2024TheIT}
Soto-Forero, D., Ackermann, S., Betbeder, M.-L., Henriet, J.: The Intelligent Tutoring System AI-VT with Case-Based Reasoning and Real Time Recommender Models. In: International Conference on Case-Based Reasoning (2024).

\bibitem{Stottler1999ACR}
Stottler, R. H., Ramachandran, S.: A Case-Based Reasoning Approach to Internet Intelligent Tutoring Systems (ITS) and ITS Authoring. In: The Florida AI Research Society (1999).

\bibitem{Soh2007IntegratedIC}
Soh, L.-K.: Integrated Introspective Case-Based Reasoning for Intelligent Tutoring Systems. In: AAAI Conference on Artificial Intelligence (2007).

\bibitem{Masood2017CasebasedRI}
Masood, M., Mokmin, N. A. M.: Case-based Reasoning Intelligent Tutoring System: An Application of Big Data and IoT. In: Proceedings of the 1st International Conference on Big Data Research (2017).

\bibitem{Hafidi2013DesignAE}
Hafidi, M., Bensebaa, T.: Design and evaluation of an adaptive and intelligent tutoring system by expert system. In: Intelligent Decision Technologies, vol. 7, pp. 253--264 (2013).

\bibitem{Zarandi2012AFE}
Zarandi, M. H. F., Khademian, M., Minaei-Bidgoli, B., Turksen, I. B.: A Fuzzy Expert System Architecture for Intelligent Tutoring Systems: A Cognitive Mapping Approach. In: Journal of Intelligent Learning Systems and Applications, vol. 04, pp. 29--40 (2012).

\bibitem{deCarvalho2020IntelligentTS}
de Carvalho, S. D., de Melo, F. R., Flôres, E. L., Pires, S. R., Loja, L. F. B.: Intelligent tutoring system using expert knowledge and Kohonen maps with automated training. In: Neural Computing and Applications, vol. 32, pp. 13577--13589 (2020).

\bibitem{Mitrovic2011ThermoTutorAI}
Mitrovic, A., Williamson, C., Bebbington, A., Mathews, M., Suraweera, P., Martin, B., Thomson, D., Holland, J.: Thermo-Tutor: An Intelligent Tutoring System for thermodynamics. In: 2011 IEEE Global Engineering Education Conference (EDUCON), pp. 378--385 (2011).

\bibitem{DMello2012GazeTA}
D’Mello, S. K., Olney, A. M., Williams, C., Hays, P.: Gaze tutor: A gaze-reactive intelligent tutoring system. In: International Journal of Human-Computer Studies, vol. 70, pp. 377--398 (2012).

\bibitem{Pereira2023ECGTA}
Pereira, L. A., Leão, L. L. S., Dermeval, D., Coelho, J. A. P. M.: ECG Tutor: a gamified intelligent tutoring system for electrocardiogram teaching. In: Revista Brasileira de Educação Médica (2023).

\bibitem{Shah2024AIPoweredPL}
Shah, T.: AI-Powered Personalised Learning Plans for Intelligent Tutoring Systems. In: International Journal for Research in Applied Science and Engineering Technology (2024).

\bibitem{Sottilare2018ExaminingCA}
Sottilare, R. A., Salas, E.: Examining Challenges and Approaches to Building Intelligent Tutoring Systems for Teams. In: Research on Managing Groups and Teams (2018).

\bibitem{Kochmar2020AutomatedPF}
Kochmar, E., Vu, D. D., Belfer, R., Gupta, V., Serban, I., Pineau, J.: Automated Personalized Feedback Improves Learning Gains in An Intelligent Tutoring System. In: Artificial Intelligence in Education, vol. 12164, pp. 140--146 (2020).

\bibitem{Lampropoulos2022AugmentedRA}
Lampropoulos, G., Keramopoulos, E., Diamantaras, K. I., Evangelidis, G.: Augmented Reality and Virtual Reality in Education: Public Perspectives, Sentiments, Attitudes, and Discourses. In: Education Sciences (2022).

\bibitem{Kulik2016EffectivenessOI}
Kulik, J. A., Fletcher, J. D.: Effectiveness of Intelligent Tutoring Systems. In: Review of Educational Research, vol. 86, pp. 42--78 (2016).

\bibitem{Chen2020ApplicationAT}
Chen, X., Xie, H., Zou, D., Hwang, G.-J.: Application and theory gaps during the rise of Artificial Intelligence in Education. In: Computers and Education: Artificial Intelligence, vol. 1, 100002 (2020).

\bibitem{BaccaAcosta2021AugmentedRI}
Bacca-Acosta, J., Ávila-Garzón, C., Kinshuk, Duarte, J., Betancourt, J.: Augmented Reality in Education: An Overview of Twenty-Five Years of Research. In: Contemporary Educational Technology, vol. 13, ep302 (2021).

\bibitem{Tafazoli2019IntelligentLT}
Tafazoli, D., Gómez María, E., Huertas Abril, C. A.: Intelligent Language Tutoring System: Integrating Intelligent Computer-Assisted Language Learning Into Language Education. In: International Journal of Information and Communication Technology Education, vol. 15, pp. 60--74 (2019).

\bibitem{Kelkar2022BetweenAA}
Kelkar, S.: Between AI and Learning Science: The Evolution and Commercialization of Intelligent Tutoring Systems. In: IEEE Annals of the History of Computing, vol. 44, pp. 20--30 (2022).

\bibitem{Deng2024ResearchOT}
Deng, Y., Wu, C.: Research on the Construction of Mobile Terminal Assisted Language Learning Model Based on Artificial Intelligence Technology. In: Journal of Intelligence and Knowledge Engineering (2024).

\bibitem{Dahbi2023IntegratingAI}
Dahbi, M.: Integrating an Intelligent Language Tutoring System in Teaching English Grammar. In: Arab World English Journal (2023).

\bibitem{Chen2022EducationTA}
Chen, X., Bear, E., Hui, B., Santhi Ponnusamy, H., Meurers, W. D.: Education Theories and AI Affordances: Design and Implementation of an Intelligent Computer Assisted Language Learning System. In: International Conference on Artificial Intelligence in Education (2022).











\end{thebibliography}

\end{document}